\documentclass[twocolumn]{aastex63}
\usepackage{appendix}
\usepackage{natbib}
\usepackage{subfigure}
\usepackage[encapsulated]{CJK}

\usepackage{amsmath} 

\newcommand{\oiii}{[\ion{O}{3}]}
\newcommand{\nii}{[\ion{N}{2}]}
\newcommand{\nev}{[\ion{Ne}{5}]}
\newcommand{\msun}{$M_{\odot}$}

\newcommand{\mbh}{$M_{\rm BH}$}

\newcommand{\jwst}{\ensuremath{JWST}}

\newcommand{\bronzenr}{40}

\newcommand{\goldnr}{17}

\newcommand{\sourcenum}{17} 

\usepackage[encapsulated]{CJK}

\begin{document}

\title{UNCOVER spectroscopy confirms a surprising ubiquity of AGN in red galaxies at $z>5$}
\author[0000-0002-5612-3427]{Jenny E. Greene}
\affiliation{Department of Astrophysical Sciences, Princeton University, 4 Ivy Lane, Princeton, NJ 08544, USA}

\author[0000-0002-2057-5376]{Ivo Labbe}
\affiliation{Centre for Astrophysics and Supercomputing, Swinburne University of Technology, Melbourne, VIC 3122, Australia}

\author[0000-0003-4700-663X]{Andy D. Goulding}
\affiliation{Department of Astrophysical Sciences, Princeton University, 4 Ivy Lane, Princeton, NJ 08544, USA}

\author[0000-0001-6278-032X]{Lukas J. Furtak}
\affiliation{Department of Physics, Ben-Gurion University of the Negev, P.O. Box 653, Be’er-Sheva 84105, Israel}

\author[0009-0009-9795-6167]{Iryna Chemerynska}
\affiliation{Institut d'Astrophysique de Paris, CNRS, Sorbonne Universit\'e, 98bis Boulevard Arago, 75014, Paris, France}

\author[0000-0002-5588-9156]{Vasily Kokorev}
\affiliation{Kapteyn Astronomical Institute, University of Groningen, 9700 AV Groningen, The Netherlands}

\author[0000-0001-8460-1564]{Pratika Dayal}\affil{Kapteyn Astronomical Institute, University of Groningen, 9700 AV Groningen, The Netherlands}

\author[0000-0003-2919-7495]{Christina C. Williams}\affiliation{NSF’s National Optical-Infrared Astronomy Research Laboratory, 950 N. Cherry Avenue, Tucson, AZ 85719, USA}\affiliation{Steward Observatory, University of Arizona, 933 North Cherry Avenue, Tucson, AZ 85721, USA}

\author[0000-0001-9269-5046]{Bingjie Wang (\begin{CJK*}{UTF8}{gbsn}王冰洁\ignorespacesafterend\end{CJK*})}
\affiliation{Department of Astronomy \& Astrophysics, The Pennsylvania State University, University Park, PA 16802, USA}
\affiliation{Institute for Computational \& Data Sciences, The Pennsylvania State University, University Park, PA 16802, USA}
\affiliation{Institute for Gravitation and the Cosmos, The Pennsylvania State University, University Park, PA 16802, USA}

\author[0000-0003-4075-7393]{David J. Setton}\thanks{Brinson Prize Fellow}\affiliation{Department of Astrophysical Sciences, Princeton University, 4 Ivy Lane, Princeton, NJ 08544, USA}

\author[0000-0002-6523-9536]{Adam J.\ Burgasser}
\affiliation{Department of Astronomy \& Astrophysics, UC San Diego, La Jolla, CA 92093, USA}

\author[0000-0001-5063-8254]{Rachel Bezanson}
\affiliation{Department of Physics and Astronomy and PITT PACC, University of Pittsburgh, Pittsburgh, PA 15260, USA}

\author[0000-0002-7570-0824]{Hakim Atek}
\affiliation{Institut d'Astrophysique de Paris, CNRS, Sorbonne Universit\'e, 98bis Boulevard Arago, 75014, Paris, France}

\author[0000-0003-2680-005X]{Gabriel Brammer} \affiliation{Cosmic Dawn Center (DAWN), Niels Bohr Institute, University of Copenhagen, Jagtvej 128, K{\o}benhavn N, DK-2200, Denmark}

\author[0000-0002-7031-2865]{Sam E. Cutler}\affiliation{Department of Astronomy, University of Massachusetts, Amherst, MA 01003, USA}

\author[0000-0002-1109-1919]{Robert Feldmann}
\affiliation{Institute for Computational Science, University of Zurich, Zurich, CH-8057, Switzerland}

\author[0000-0001-7201-5066]{Seiji Fujimoto}\altaffiliation{Hubble Fellow}
\affiliation{Department of Astronomy, The University of Texas at Austin, Austin, TX 78712, USA}

\author[0000-0002-3254-9044]{Karl Glazebrook}\affiliation{Centre for Astrophysics and Supercomputing, Swinburne University of Technology, PO Box 218, Hawthorn, VIC 3122, Australia}

\author[0000-0002-2380-9801]{Anna de Graaff}
\affiliation{Max-Planck-Institut f{\"u}r Astronomie, K{\"o}nigstuhl 17, D-69117, Heidelberg, Germany}

\author[0000-0002-3475-7648]{Gourav Khullar}
\affiliation{Department of Physics and Astronomy and PITT PACC, University of Pittsburgh, Pittsburgh, PA 15260, USA}

\author[0000-0001-6755-1315]{Joel Leja}
\affiliation{Department of Astronomy \& Astrophysics, The Pennsylvania State University, University Park, PA 16802, USA}
\affiliation{Institute for Computational \& Data Sciences, The Pennsylvania State University, University Park, PA 16802, USA}
\affiliation{Institute for Gravitation and the Cosmos, The Pennsylvania State University, University Park, PA 16802, USA}

\author[0000-0001-9002-3502]{Danilo Marchesini}
\affiliation{Department of Physics \& Astronomy, Tufts University, MA 02155, USA}

\author[0000-0003-0695-4414]{Michael V. Maseda}
\affiliation{Department of Astronomy, University of Wisconsin-Madison, 475 N. Charter St., Madison, WI 53706 USA}

\author[0000-0003-2871-127X]{Jorryt Matthee}
\affiliation{Department of Physics, ETH Zurich, Wolfgang-Pauli-Strasse 27, 8093 Zurich, Switzerland}
\affiliation{Institute of Science and Technology Austria (IST Austria), Am Campus 1, Klosterneuburg, Austria}

\author[0000-0001-8367-6265]{Tim B. Miller}
\affiliation{Department of Astronomy, Yale University, New Haven, CT 06511, USA}
\affiliation{Center for Interdisciplinary Exploration and Research in Astrophysics (CIERA) and Department of Physics \& Astronomy, Northwestern University, IL 60201, USA}

\author[0000-0003-3997-5705]{Rohan~P.~Naidu}
\altaffiliation{NASA Hubble Fellow}
\affiliation{MIT Kavli Institute for Astrophysics and Space Research, 77 Massachusetts Ave., Cambridge, MA 02139, USA}

\author[0000-0003-2804-0648 ]{Themiya Nanayakkara}
\affiliation{Centre for Astrophysics and Supercomputing, Swinburne University of Technology, PO Box 218, Hawthorn, VIC 3122, Australia}

\author[0000-0001-5851-6649]{Pascal A. Oesch}
\affiliation{Department of Astronomy, University of Geneva, Chemin Pegasi 51, 1290 Versoix, Switzerland}
\affiliation{Cosmic Dawn Center (DAWN), Niels Bohr Institute, University of Copenhagen, Jagtvej 128, K{\o}benhavn N, DK-2200, Denmark}

\author[0000-0002-9651-5716]{Richard Pan}\affiliation{Department of Physics and Astronomy, Tufts University, 574 Boston Ave., Medford, MA 02155, USA}

\author[0000-0001-7503-8482]{Casey Papovich}
\affiliation{Department of Physics and Astronomy, Texas A\&M University, College
Station, TX, 77843-4242 USA}
\affiliation{George P.\ and Cynthia Woods Mitchell Institute for
 Fundamental Physics and Astronomy, Texas A\&M University, College
 Station, TX, 77843-4242 USA}

\author[0000-0002-0108-4176]{Sedona H. Price}\affiliation{Department of Physics and Astronomy and PITT PACC, University of Pittsburgh, Pittsburgh, PA 15260, USA}

\author[0000-0002-8282-9888]{Pieter van Dokkum}
\affiliation{Astronomy Department, Yale University, 52 Hillhouse Ave,
New Haven, CT 06511, USA}

\author[0000-0003-1614-196X]{John R. Weaver}
\affiliation{Department of Astronomy, University of Massachusetts, Amherst, MA 01003, USA}

\author[0000-0001-7160-3632]{Katherine E. Whitaker} \affiliation{Department of Astronomy, University of Massachusetts, Amherst, MA 01003, USA}\affiliation{Cosmic Dawn Center (DAWN), Denmark}

\author[0000-0002-0350-4488]{Adi Zitrin}\affiliation{Department of Physics, Ben-Gurion University of the Negev, P.O. Box 653, Be’er-Sheva 84105, Israel}

\date{September 2023}

\begin{abstract}
\jwst\ is revealing a new population of dust-reddened broad-line active galactic nuclei (AGN) at redshifts $z\gtrsim5$. Here we present deep NIRSpec/Prism spectroscopy from the Cycle 1 Treasury program UNCOVER of 17 AGN candidates selected to be compact, with red continua in the rest-frame optical but with blue slopes in the UV. From NIRCam photometry alone, they could have been dominated by dusty star formation or AGN. Here we show that the majority of the compact red sources in UNCOVER are dust-reddened AGN: 60\% show definitive evidence for broad-line H$\alpha$ with FWHM$>2000$~km/s, for 20\% current data are inconclusive, and $20\%$ are brown dwarf stars. We propose an updated photometric criterion to select red $z>5$ AGN that excludes brown dwarfs and is expected to yield $>80\%$ AGN. Remarkably, among all $z_{\rm phot}>5$ galaxies with F277W$-$F444W$>1$ in UNCOVER at least $33\%$ are AGN regardless of compactness, climbing to at least $80$\% AGN for sources with F277W$-$F444W$>1.6$. 
The confirmed AGN have black hole masses of $10^7-10^9$~\msun. While their UV-luminosities ($-16>M_{\rm UV}>-20$~AB mag) are low compared to UV-selected AGN at these epochs, consistent with percent-level scattered AGN light or low levels of unobscured star formation, the inferred bolometric luminosities are typical of $10^7-10^9$~\msun\ black holes radiating at $\sim 10-40\%$ of Eddington. The number densities are surprisingly high at $\sim10^{-5}$~Mpc$^{-3}$~mag$^{-1}$, 100 times more common than the faintest UV-selected quasars, while accounting for $\sim1\%$ of the UV-selected galaxies. While their UV-faintness suggest they may not contribute strongly to reionization, their ubiquity poses challenges to models of black hole growth. 
\end{abstract}


\section{Introduction}
\label{sec:intro}

Over the past decade, large-area surveys have discovered hundreds of UV-luminous active galactic nuclei (AGN) at $z>5$  \citep[e.g.,][]{Mortlock:2011,Banados:2018,Fan:2019,Wang:2021,Harikane:2022,Fan:2023}. Unlike lower redshifts, the number densities of UV-selected AGN at $z>5$ are not strongly luminosity dependent for $M_{\rm UV}$ fainter than $\sim -25$~mag \citep{Matsuoka:2018,Matsuoka:2023}, while the galaxy LF rises steeply, so that UV-selected AGN fainter than about $M_{\rm UV} \approx -25$~mag make up $<0.1$\% of the galaxy population at high redshift. Determining whether or not black hole growth is preceding or lagging galaxy growth at these epochs has important implications for the seeding and co-evolution of black holes and galaxies \citep[e.g.,][]{VolonteriReines:2016,Dayal:2019, Inayoshi:2020,Greene:2020, Zhang:2023}, for the sources of reionization \citep[e.g.,][]{Madau:2015, Dayal:2020,Trebitsch:2023}, and potentially for the sources of gravitational waves \citep[e.g.,][]{AmaraSeoane:2023,Somalwar:2023}.

With the advent of the James Webb Space Telescope \citep[\jwst][]{Gardner:2023}, we have begun to identify the heretofore missing UV-faint AGN. They have been discovered through broad Balmer lines \citep{Kocevski:2023,Harikane:2023,Oesch:2023,Larson:2023,Ubler:2023,Barro:2023,Matthee:2023,Maiolino:2023}, from color and morphology \citep{Onoue:2023, Ono:2022, Furtak:2022, Endsley:2022,Hainline:2023,Leung:2023,Yang:2023}, and from X-ray emission \citep{Bogdan:2023,Goulding:2023}. The number densities of these \jwst-selected sources is roughly a few percent of the galaxy population, and while they are UV faint, their implied bolometric luminosities span a broad range ($L_{\rm bol} \sim 10^{43}-10^{46}$~erg/s), implying a wide range of \mbh$\sim 10^6-10^9$~\msun. 

An interesting sub-component of the \jwst-selected AGN population are quite red (e.g., F2777$-$F444W>1) \citep[see also][]{Fujimoto:2022,Furtak:2022}. For instance, \citet{Matthee:2023} spectroscopically identified a sample of broad-line AGN that all appear as ``little red dots'' with steep red continua in the rest-frame optical \citep[see also][]{Kocevski:2023}. Other \jwst+broad-line selected samples show large red fractions of 10-20\% \citep{Harikane:2023,Maiolino:2023}. At $z<2$, reddened broad-line AGN are known \citep[e.g.,][]{Glikman:2012,Banerji:2015}, but these particular spectral energy distributions, with both a steep red optical continuum and an additional UV component are quite rare \citep[e.g.,][]{Noboriguchi:2019}. 

What has not been clear to date is whether a photometrically selected sample of compact red sources with a significant UV component are dominated by AGN as well, or whether they may be powered by star formation \citep{Barro:2023,Akins:2023,Maiolino:2023}. Recently, we \citep[][L23 hereafter]{Labbe:2023uncover} published a large photometric sample of compact red sources from the Cycle 1 \jwst\ program UNCOVER \citep{Bezanson:2022}. We have now obtained follow-up NIRSpec spectroscopy for 15 of the \bronzenr\ galaxies presented in L23. Here, we explore the nature of the sample, and show that our selection indeed provides a very high yield of $z>4$ AGN.

We present the UNCOVER data in \S \ref{sec:data}, and review the photometric selection in \S \ref{sec:sample}. The spectroscopic analysis, and in particular the identification of broad H$\alpha$ emission lines, is presented in \S \ref{sec:findbroad}. Throughout, we assume a concordance cosmology with H$_0$=70, $\Omega_{\Lambda} = 0.7$, $\Omega_{M} = 0.3$ \citep{Hinshaw:2013}.

\section{Data, Sample, and Spectroscopic Follow-up}
\label{sec:data}

In this section we briefly describe the UNCOVER survey \citep[\S \ref{sec:uncover}][]{Bezanson:2022} and the MSA/PRISM spectroscopy and reductions \ref{sec:prism}. Many exciting results have already come from the PRISM spectroscopy of the compact red sources, including a triply-imaged $z=7$ AGN \citep{Furtak:2023nature}, a broad-line AGN at z=8.5 \citep{Kokorev:2023,Fujimoto:2023overdense}, and three brown dwarfs which can show similar SEDs as red AGN \citep{Langeroodi:2023,Burgasser:2023}. We have also reported two $z>12$ galaxies \citep{Wang:2023highz}, and some of the faintest known targets in the epoch of reionization \citep{Atek:2023eor}.

\subsection{UNCOVER Photometry}
\label{sec:uncover}

Our search is performed using the \emph{JWST} Cycle 1 Treasury program Ultradeep NIRSpec and NIRCam ObserVations before the Epoch of Reionization \citep[UNCOVER;][]{Bezanson:2022}. UNCOVER imaging was completed in November 2022, comprising ultradeep ($29-30$~AB mag) imaging over 45 arcmin$^2$ in the galaxy cluster Abell 2744. This well-studied Frontier Field cluster \citep{Lotz:2017} at $z=0.308$ has one of the largest high-magnification areas of known clusters, and thus made an excellent target for deep (4-6 hours per filter) imaging across seven NIRCam filters (F115W, F150W, F200W, F277W, F356W, F410M, F444W). The nominal depth of $\sim 30$~mag can comfortably reach sources as faint as 31.5~mag with the help of magnification.  Photometric catalogs \citep{Weaver:2023} including existing \emph{HST} data have been made available to the public, and the lens model is also publicly available \citep{Furtak:2022b_SLmodel}. The initial selection of objects is based on the UNCOVER Data Release DR1 images and catalogs \citep{Weaver:2023}. 

\citet{Fujimoto:2023} present deep ALMA 1.2~mm continuum imaging in Abell 2744. A wider, deeper 1.2~mm map of the full NIRCam UNCOVER area was newly obtained in Cycle 9 (\#2022.1.00073.S; S. Fujimoto in prep), 
reaching continuum r.m.s. sensitivity of $33 \rm{\mu Jy}$ in the deepest areas. Prior-based photometry is extracted for all sources by measuring the ALMA flux in the natural resolution map (beam $\approx0.7-0.8\arcsec$) at the NIRCam positions.

The UNCOVER region additionally has full coverage with high spatial resolution X-ray imaging using the {\it Chandra} ACIS-I detector, which upon completion will be at $\sim$2.3~Ms depth (PI:A.Bogdan). This X-ray data has already identified the highest-redshift X-ray AGN to date, UHZ-1, spectroscopically confirmed at $z=10.1$ \citep{Bogdan:2023,Goulding:2023}. There is also deep Jansky Very Large Array data from program 22A-017 (PI Murphy) that may provide complementary radio imaging.

\begin{figure*}
\hspace{-2mm}
\includegraphics[width=0.45\textwidth]{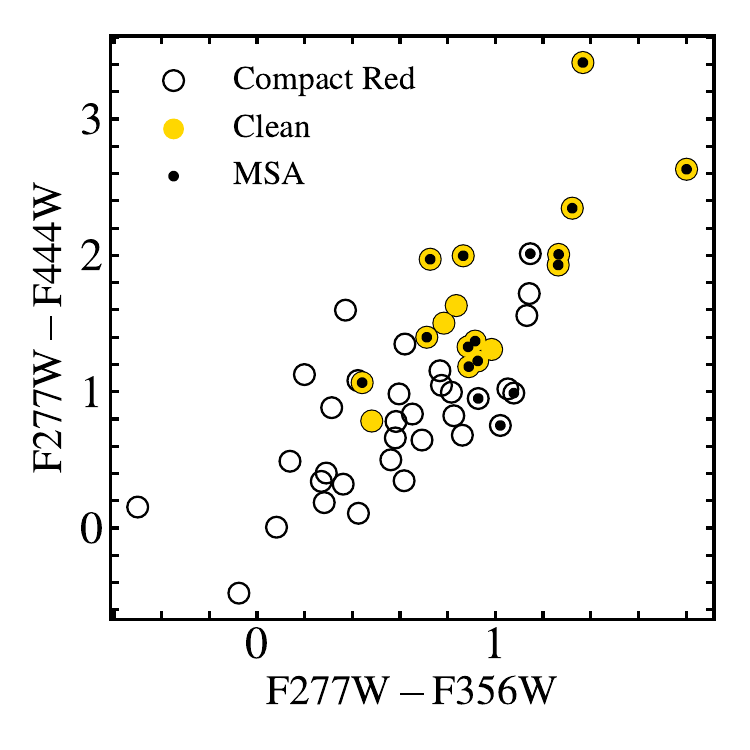}
\includegraphics[width=0.45\textwidth]{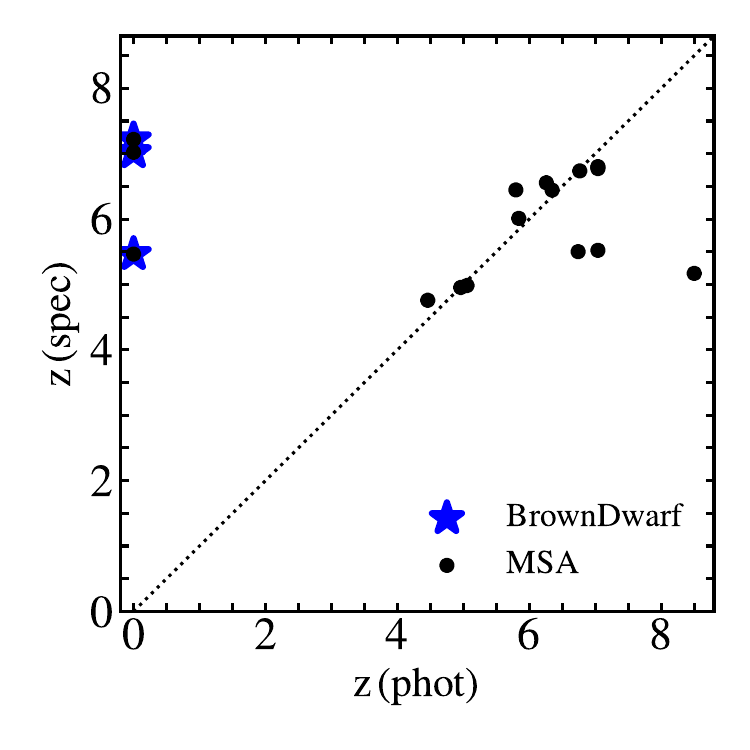}
\caption{Left: The primary color-color selection used to select the compact red sources. We show the full sample of 40 compact red sources (open circles), the 17 clean targets (yellow) and the spectroscopically targeted sources (black dots are those observed with the MSA). Comparison of the photometric redshifts measured from the NIRCam photometry using custom templates from L23, as compared with the spectroscopic redshifts. The brown dwarfs are indicated with stars.
}
\label{fig:target}
\end{figure*}

\subsection{PRISM Spectroscopy and Reductions}
\label{sec:prism}

\subsubsection{MSA Observational Setup}

All \sourcenum\ spectroscopic targets were observed with the Micro-Shutter Assembly (MSA) follow-up program of the UNCOVER \jwst\ field Abell 2744 \citep{Bezanson:2022}. The UNCOVER NIRSpec/PRISM observations were taken over 7 MSA configurations. These observations employed a 2-POINT-WITH-NIRCam-SIZE2
dither pattern and a 3 shutter slit-let nod pattern at an aperture angle of $\sim 44$ degrees. The observational design for the photometric component is described in detail in \citet{Bezanson:2022}, the catalog is described by \citet{Weaver:2023}, the photometric redshifts are explored in some depth by \citet{Wang:2023}, and the spectroscopic experimental design and reductions are explained by Price et al. (2023, in prep).

\subsubsection{NIRSpec/PRISM Data Reduction}

The data reduction is performed with \texttt{msaexp} \citep[v0.6.10,][]{Brammer:22}, beginning from the level 2 products downloaded from MAST\footnote{Available from: \url{http://dx.doi.org/10.17909/8k5c-xr27}}. \texttt{msaexp} then corrects for 1/f noise, masks snowballs, and removes the bias frame-by-frame. The \jwst\ reduction pipeline is used to apply WCS, to identify each slit, and then to perform flat-fielding and to apply photometric correction. 2D slits are extracted and drizzled onto a common grid to make vertically shifted and stacked 2D spectra, to which a local background subtraction is applied. The optimal extraction uses a Gaussian model on the collapsed spectrum with a free center and width \citep[e.g.,][]{Horne:1986}. To flux calibrate the spectra, the single-mask extracted 1D spectra are convolved with the broad/medium band filters, and then we compare to the total photometry (\citealt{Weaver:2023}), and model the wavelength dependent linear correction with a first order polynomial. Reduced data will be presented in Price et al. (2023, in prep). 

\subsection{Gravitational magnification} \label{sec:SL}

Throughout this study we use the latest version (\texttt{v1.1}) of the \citet{Furtak:2022b_SLmodel} strong lensing model of Abell~2744\footnote{The \texttt{v1.1} lensing maps are publicly available at \url{https://jwst-uncover.github.io/DR1.html\#LensingMaps}}. The parametric strong lensing model was constructed with an updated version of the \citet{Zitrin:2015} parameteric method \citep{Pascale:2022,Furtak:2022b_SLmodel} and includes 421 cluster member galaxies and five smooth cluster-scale dark matter halos. The \texttt{v1.1} has been updated with 5 new multiple image systems and and additional spectroscopic redshifts \citep{Bergamini:2023_GLASS}, and is thus constrained with 141 multiple images belonging to 48 sources across the several main clumps of the cluster. The final lens plane image reproduction error is $\Delta_{\mathrm{RMS}}=0.51\arcsec$. We compute magnifications and their uncertainties for our sample at each object's position and spectroscopic redshift.

\section{Photometric and Spectroscopic Sample}
\label{sec:sample}

First, we briefly review the basic properties of the photometrically selected AGN sample (\S \ref{sec:parent}, L23), then we summarize the objects targeted for MSA/PRISM spectroscopy.

\subsection{Sample of Red Compact Sources}
\label{sec:parent}

We build the photometric sample of AGN candidates in a number of key steps. First we select compact and red sources, with the following color and morphology cuts.  With $SNR(\rm{F444W}) > 14 \ \& \ m_{\rm F444W} < 27.7$~mag within a 0.32\arcsec\ aperture, we select sources that are ({\it red1} $\vert$ {\it red2} ) \& {\it compact}, where 

$${\it red1} = (\rm{F115W-F150W} < 0.8) \, \land $$
$$  (\rm{F200W-F277W} > 0.7) \, \land $$ 
$$ (\rm{F200W-F356W} > 1.0) \,  $$  

$${\it red2} = (\rm{F150W-F200W} < 0.8)  \, \land $$ 
$$ (\rm{F277W-F356W} > 0.7) \, \land $$
$$ (\rm{F277W-F444W} > 1.0) \,  $$

And:
$$ {\it compact} = f_{\rm{F444W}}(0.4\arcsec) / f_{\rm{F444W}}(0.2\arcsec) < 1.7 $$ 

for $f_{\rm{F444W}}(0.4\arcsec)$ measured within a 0.4\arcsec\ diameter aperture.
This initial sample comprises \bronzenr\ sources, the ``compact red'' sample. We show the full selection in Figure \ref{fig:target}; the main difference between $red1$ and $red2$ is that $red2$ favors higher redshift galaxies owing to the redder filters used. The $red2$ sample accounts for $58\%$ of the photometric sample and has been the focus for spectroscopic follow-up, representing $82\%$  of spectroscopic targets.

For context, L23 color criteria, specifically $red2$, to select the compact red sample are similar to those used by \citet{Labbe:2023} to select ``v-shaped'' SEDs as candidates for massive galaxies at similar epochs, with an important distinction. The $red2$ applies a more stringent set of cuts to ensure that the sources are point-source dominated and show two red consecutive colors in the LW NIRCam filters, thus favoring sources with red continuum slopes rather than contribution of emission lines or a continuum break. Overlap betweent the samples is discussed in \S 5.  

As a final step, L23 identified a ``clean'' sample of \goldnr\ objects keeping only sources where two-dimensional image fitting in F356W indicated $<$50\% of the light residing in an extended component and where spectral energy distribution (SED) fitting indicated the broad-band SEDs could not be fit without an AGN component. 

In nearly all cases, the highest priority targets are those with deep ALMA limits, because in these cases we can effectively rule out dusty star formation as the origin of the red continuum, unless we were to invoke much hotter dust than is seen at these (or any) redshift. However, only 3/20 targets with ALMA coverage favored SED solutions without an AGN component, so we will assume that the NIRCam color selected sample is largely representative of the SED selected sample. We will discuss the AGN yield under different cuts in \S \ref{sec:yield}.

\begin{deluxetable*}{ccccccccccccc}
\tabletypesize{\footnotesize}
\tablecolumns{13}
\tablewidth{0pt}
\tablecaption{ Sample \label{table:thesample}}
\tablehead{
\colhead{MSAID} & \colhead{RA} & \colhead{Dec} & \colhead{$z_{\rm spec}$} & \colhead{$\mu$} & \colhead{$\mu_{\rm low}$} & \colhead{$\mu_{\rm high}$} & \colhead{F444W} & \colhead{277-444} & \colhead{277-356}  & \colhead{Flg} & \colhead{MSA} & \colhead{T$_{\rm exp}$} \\ 
\colhead{(1)} & \colhead{(2)} & \colhead{(3)} & \colhead{(4)} & \colhead{(5)} & \colhead{(6)} & \colhead{(7)} & \colhead{(8)} & \colhead{(9)} & \colhead{(10)} & \colhead{(11)} & \colhead{(12)} & \colhead{(13)}
}
\startdata
2008 & 3.592423 & $-$30.432828 & 6.74$\pm$0.00 & 1.69 & 1.68 & 1.72 & 27.3 & 1.39 & 1.06 & 0 & 1 & 2.7\\
 4286 & 3.619202 & $-$30.423270 & 5.84$\pm$0.00 & 1.62 & 1.61 & 1.64 & 24.8 & 2.03 & 1.19 & 0 &  2& 2.7 \\
 10686 & 3.550838 & $-$30.406598 & 5.05$\pm$0.00 & 1.44 & 1.45 & 1.47 & 24.3 & 1.05 & 0.49 & 1 & 1  &2.7 \\
 13123$^a$ & 3.579829 & $-$30.401570 & 7.04$\pm$0.00 & 6.15 & 5.96 & 6.69 & 25.0 & 2.66 & 1.89 & 1 & 2,3,5,6,7  &17.4 \\
 13821 & 3.620607 & $-$30.399951 & 6.34$\pm$0.00 & 1.59 & 1.57 & 1.61 & 25.0 & 2.33 & 1.40 & 1  &1 &2.7 \\
 15383$^a$ & 3.583534 & $-$30.396678 & 7.04$\pm$0.00 & 7.29 & 5.78 & 7.30 & 25.3 & 2.45 & 1.71 & 1 &2,3,6 &9.9\\
 16594$^a$ & 3.597203 & $-$30.394330 & 7.04$\pm$0.00 & 3.55 & 3.38 & 3.70 & 26.3 & 1.97 & 1.26 & 1 & 1,5,6,7& 14.7\\
 20466 & 3.640409 & $-$30.386437 & 8.50$\pm$0.00 & 1.33 & 1.31 & 1.34 & 26.2 & 1.92 & 0.72 & 1 &2 &2.7 \\
 23608 & 3.542815 & $-$30.380646 & 5.80$\pm$0.00 & 2.07 & 2.07 & 2.13 & 24.9 & 0.79 & 0.88 & 0 &3& 2.7 \\
 28876 & 3.569596 & $-$30.373222 & 7.04$\pm$0.00 & 2.70 & 2.60 & 2.73 & 26.8 & 2.10 & 1.49 & 1 &1,4 &6.4 \\
 32265 & 3.537530 & $-$30.370168 & 0.00$\pm$0.00 & \nodata & \nodata & \nodata & 27.3 & 0.97 & 0.68 & 1 &3,5,6,7& 14.7 \\
 33437 & 3.546419 & $-$30.366245 & 0.00$\pm$0.00 & \nodata & \nodata & \nodata & 27.0 & 2.06 & 0.97 & 1 &3,5,6,7& 14.7 \\
 35488 & 3.578984 & $-$30.362598 & 6.26$\pm$0.00 & 3.38 & 3.14 & 3.74 & 24.5 & 0.83 & 0.99 & 0 &1& 2.7 \\
 38108 & 3.530009 & $-$30.358013 & 4.96$\pm$0.00 & 1.59 & 1.58 & 1.62 & 24.7 & 1.06 & 0.83 & 1 &4 &3.7 \\
 39243 & 3.513894 & $-$30.356024 & 0.00$\pm$0.00 & \nodata & \nodata & \nodata & 25.6 & 3.63 & 1.53 & 1 &4 &3.7 \\
 41225 & 3.533994 & $-$30.353308 & 6.76$\pm$0.00 & 1.50 & 1.49 & 1.53 & 25.9 & 1.13 & 0.71 & 1 &4& 3.7 \\
 45924 & 3.584758 & $-$30.343630 & 4.46$\pm$0.00 & 1.59 & 1.58 & 1.65 & 22.1 & 0.54 & 0.94 & 1 &4,5,6,7& 15.7\\
\enddata
\tablecomments{Table of objects that satisfy $(red1 \vert red2)$ and $compact$. $a$ are the three images of the multiply-imaged red AGN presented in \citet{Furtak:2022,Furtak:2023nature}. 
Column (1): MSA ID.
Column (2): R.A. 
Column (3): Dec.
Column (4): Spectroscopic redshifts {\bf }
Column (5): Total magnification ($\mu$) based on the \texttt{v1.1} UNCOVER strong lensing model (see section~\ref{sec:SL}). 
Column (6): Magnification 68\% low value. 
Column (7): Magnification 68\% high value. 
Column (8): F444W mag.
Column (9): F277W-F444W color (mag). 
Column (10): F277W-F356W color (mag).
Column (11): Flag for the high-priority photometric sample. 
Column (12): MSA number (1-7).
Column (13): Total exposure time (h). 
}
\end{deluxetable*}

\subsection{Spectroscopic sample}

In designing the MSA PRISM masks, we prioritized observing the compact red sources from L23. We obtained spectra for most (75\%) of the highest-priority targets, as is shown in Figure \ref{fig:target}.

\begin{figure*}
\vspace{-2mm}
\hspace{-5mm}
\includegraphics[width=0.99\linewidth]{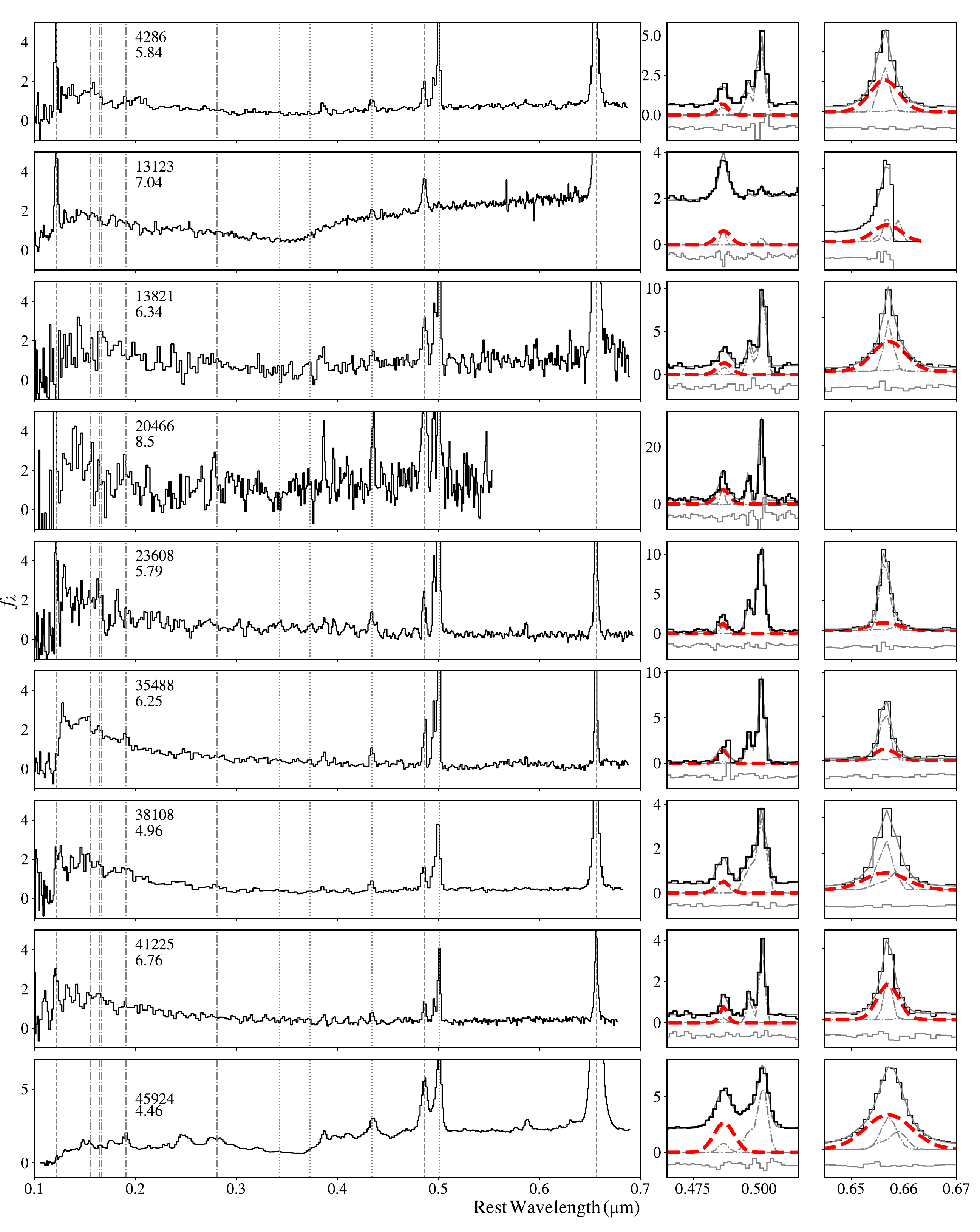}
\vskip -1mm
\caption{Left: NIRSpec/PRISM spectra of the nine confirmed broad-line AGN in the sample. Spectra are plotted in the rest wavelength and have been normalized at 2500\AA. Vertical lines indicate rest wavelengths of hydrogen lines (Ly$\alpha$, H$\beta$, H$\alpha$) in dashed, broad permitted metal lines (CIV$\lambda 1550$~\AA, HeII$\lambda 1640$~\AA, OIII]$\lambda 1663$, CIII]$\lambda 1909$~\AA, MgII$\lambda 2800$~\AA) in dash-dot, and forbidden lines (\nev$\lambda 3426$~\AA, \oiii$\lambda 3727$~\AA, \oiii$\lambda 4363$~\AA [note this is blended with H$\gamma$], \oiii$\lambda 5007$~\AA) in dotted. In the case of MSAID13123, we are actually plotting the coadded spectrum across all three images from \citet{Furtak:2023nature}.
Middle: Fits to the H$\beta$+\oiii\ spectral region. We show the data (black histogram), the full model (grey solid), the narrow-line fits (dot-dash), and the broad-line fits (thick red dashed). 
Right: Fits to the H$\alpha$+\nii region, lines as in the H$\beta$ region. We only require a significant \nii\ component in a couple of cases. } 
\label{fig:specbroad}
\end{figure*}

\begin{figure*}
\hspace{0mm}
\includegraphics[width=0.95\linewidth]{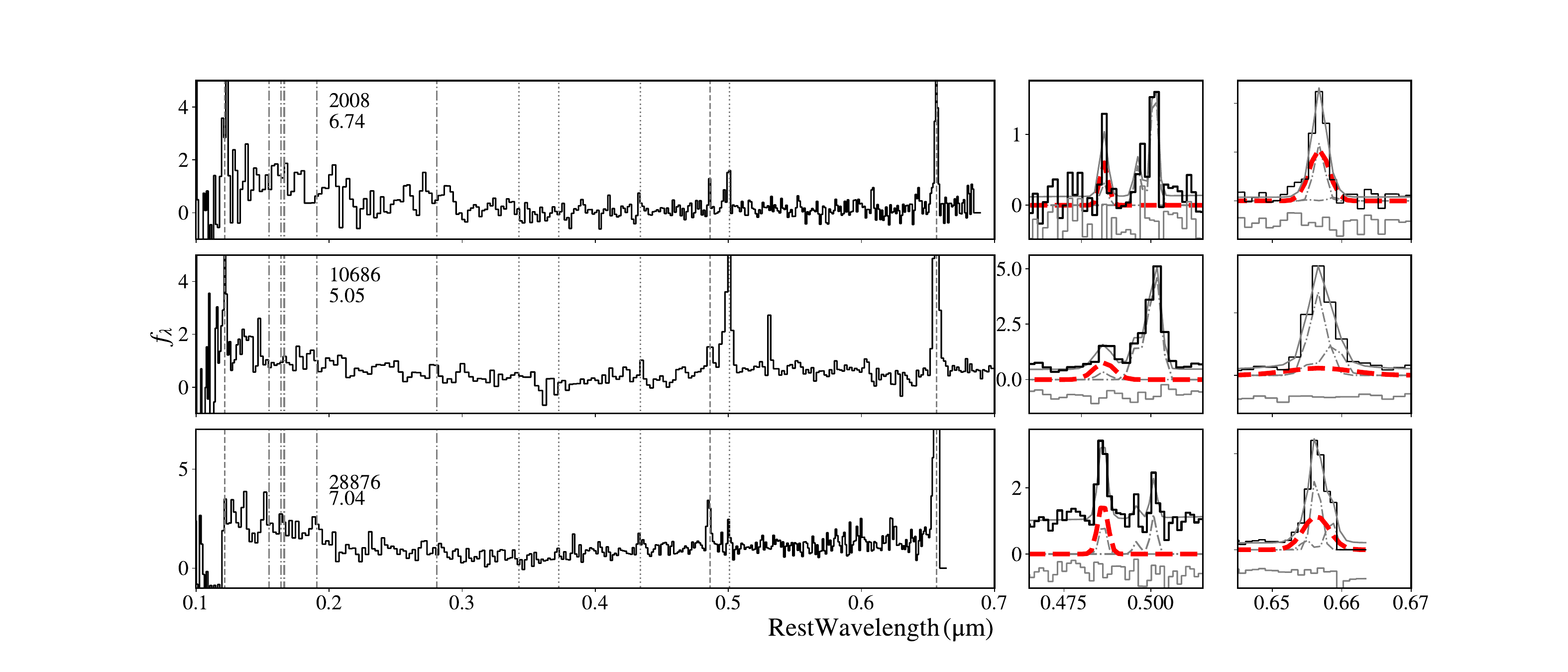}
\caption{Left: NIRSpec/PRISM spectra of the unconfirmed broad-line AGN in the sample. Spectra are plotted in the rest frame and have been normalized at 2500\AA. Vertical lines indicate rest wavelengths of hydrogen lines (Ly$\alpha$, H$\beta$, H$\alpha$) in dashed, broad permitted metal lines (CIV$\lambda 1550$~\AA, HeII$\lambda 1640$~\AA, OIII]$\lambda 1663$, CIII]$\lambda 1909$~\AA, MgII$\lambda 2800$~\AA) in dash-dot, and forbidden lines (\oiii$\lambda 3727$~\AA, \oiii$\lambda 4363$~\AA [note this is blended with H$\gamma$], \oiii$\lambda 5007$~\AA) in dotted.
Middle: Fits to the H$\beta$+\oiii\ spectral region. We show the data (black histogram), the full model (grey solid), the narrow-line fits (dot-dash), and the broad-line fits (thick red dashed). 
Right: Fits to the H$\alpha$+\nii region, lines as in the H$\beta$ region. Note that in the case of MSAID13123, the redshift of $z=7.04$ truncates the H$\alpha$ line.} 
\label{fig:specmaybe}
\end{figure*}

\section{Spectral Analysis}
\label{sec:findbroad}

We targeted with the NIRSpec/PRISM \sourcenum\ sources that were photometrically identified in the UNCOVER photometry (L23). Of these, 14 are extragalactic sources and three turn out to be cool brown dwarfs \citep{Burgasser:2023}. Three of the extragalactic sources are confirmed as multiply lensed images of the same source, A2744-QSO1 \citep{Furtak:2022,Furtak:2023nature}, leaving 12 unique non-stellar sources. All have redshifts $4.5<z_{\rm spec}<8.5$. Overall the photometric redshifts were accurate (Fig.\ \ref{fig:target}), with a median offset in $\delta z$/(1+z) of $\langle z \rangle =0.008$ and $\sigma_{MAD}=0.03$, but for some objects the redshift was underestimated \citep[e.g.,][]{Kokorev:2023}, due to degeneracy between redshifts where [OIII] or H$\alpha$ are in the F410M or F444W band.

The main goal of this paper is to determine the nature of these sources. Based on their unique properties -- detected UV continuum, red optical continuum, ALMA-undetected, and spatially compact -- we proposed based on an SED fitting that the sources are likely to be AGN. We now ask whether the spectra provide evidence for an accreting black hole. The main evidence we will present is in the form of broad Balmer lines, which has long been an accepted signature of gas orbiting the central black hole in the broad-line region \citep[e.g.,][]{Osterbrock:1977}.

\subsection{Line fitting}

From our broad wavelength coverage, from $1-5 \, \micron$, we cover from Ly$\alpha$ to Balmer lines at $4.5 < z < 8.5$. The spectra are very rich, albeit at relatively low ($R = 100-500$) dispersion. In this section, we will focus only on fitting the strong optical emission lines H$\beta$, \oiii$\lambda\lambda 4959, \, 5007$, H$\alpha$, \nii$\lambda\lambda 6584, 6548$. Note that because H$\alpha$ falls at the reddest end of the spectrum, we have enough spectral resolution to model the \nii\ lines separately from H$\alpha$.

In all cases we model the narrow forbidden lines with a single velocity width, and the ratios of the \oiii\ and \nii doublets are fixed to $\sim 3$. We model the continuum as a power-law, normalized at 5100\AA. We then perform two different fits to the H$\beta$+\oiii+H$\alpha$+\nii complex. In all fits, the radial velocities of all lines are tied together. First, we perform a {\it narrow-line} only fit, where all the lines are fixed to the same velocity width, with a flat broad prior on the width up to 800 km/s. Then we perform a {\it two-component} fit with a narrow and broad component to the Balmer lines. In this fit, all narrow lines are fixed in velocity width and fit over a narrow prior range (50-150 km/s). The ratio of narrow H$\alpha$/H$\beta$ flux is fit over a prior range of 3-20 ($A_V \sim 0-5$) under the assumption of case B recombination \citep{Osterbrock:1989}. Broad H$\beta$ and H$\alpha$ fluxes are fitted independently, and the width of H$\alpha$ is constrained to fall within a factor of two of H$\beta$. In the first fit we constrain eight free parameters (including a power-law continuum) and in the second model we add four additional free parameters.

Each model is then forward modeled through the instrument before calculating the likelihood. We use the predicted pre-launch instrumental broadening from JDOX \citep{Jakobsen:2022}, but because our targets are by definition point sources, the resolution is considerably better than the nominal resolution for a uniformly illuminated slit \citep[by up to a factor 2;][]{deGraaff2023}. On the other hand, the rectification and combination of the spectra results in additional broadening due to the relatively large pixel size compared to the instrument point spread function. We therefore increase the nominal resolution by a conservative, uniform factor of 1.3, but caution that we then neglect any wavelength-dependence of the correction factor to the line spread function. The effect on the inferred broad-line widths is negligible, but the inferred widths for the narrow-line components suffer from a systematic uncertainty. For the fitting, we define a variable wavelength grid that oversamples the native resolution by a factor of four, convolve the model with the instrumental broadening, and then resample onto the wavelength grid while preserving flux. 

The full set of narrow-only and two-component fits are presented in Figures \ref{fig:specbroad} and \ref{fig:specmaybe}, continuum fits are presented in Table \ref{table:continuum}, and broad H$\alpha$ parameters are presented in Table \ref{table:thefits}. To explore the impact of the degeneracy between broad and narrow lines, we perform an additional two-component fit in which we fix the Balmer decrement to fall between H$\alpha$/H$\beta = 4-6$, corresponding to the continuum-derived $A_V \approx 1.5$ mag as is seen in our more conservative continuum fits (\S \ref{sec:reddening}). We find that in all cases but one, the derived H$\alpha$ width agrees within 5-10\% of our fiducial fits. In the case of MSAID23608, which has a very strong narrow-line component, the broad H$\alpha$ width decreases to 1500 km/s. Thus, we are confident that overall, our broad-line widths are robust to degeneracy with the narrow components.

\begin{deluxetable*}{llrllcc}
\tabletypesize{\footnotesize}
\tablecolumns{7}
\tablewidth{0pt}
\tablecaption{Continuum Fits \label{table:continuum}}
\tablehead{
\colhead{MSAID} &  \colhead{$\alpha_{\rm opt}$} & \colhead{$\beta_{\rm UV}$} & \colhead{$A_{\rm V, 0.45}$} & \colhead{$A_{\rm V, fit}$} & 
\colhead{$L_{5100, c}$} & \colhead{$L_{\rm 5100, H\alpha}$} \\
\colhead{(1)} & \colhead{(2)} & \colhead{(3)} & \colhead{(4)} & \colhead{(5)} & \colhead{(6)} & \colhead{(7)}
}
\startdata
 2008 & 1.4$\pm$0.3 & $-$1.9$\pm$0.3 & 2.1$\pm$0.5 & 2.4$\pm$3.0 & 43.5 & 43.7 \\
 4286 & 0.9$\pm$0.1 & $-$1.6$\pm$0.1 & 1.3$\pm$0.2 & 2.7$\pm$0.2 & 44.6 & 44.7 \\
 10686 & 0.4$\pm$0.3 & $-$1.6$\pm$0.1 & 2.0$\pm$1.1 & 3.2$\pm$0.2 & 44.9 & 44.3 \\
 13123$^a$ &  1.2$\pm$0.3 & $-1.4 \pm 0.1$ & $3 \pm 0.5$ & \nodata & 44.4 & 43.8 \\
 13821 & 1.0$\pm$0.3 & $-$1.2$\pm$0.1 & 1.1$\pm$0.3 & 1.7$\pm$0.2 & 44.2 & 44.5 \\
 20466$^b$ & 0.5$\pm$0.3 & $-$0.7$\pm$0.2 & $2.1^{+1.1}_{-1}$ & $1.9 \pm 0.2$ & 44.5 & 44.9 \\
 23608 & 0.0$\pm$0.1 & $-$1.5$\pm$0.1 & 0.1$\pm$0.1 & 1.9$\pm$0.7 & 43.5 & 43.5 \\
 28876 & 1.0$\pm$0.2 & $-$2.1$\pm$0.1 & 1.4$\pm$0.1 & 2.9$\pm$1.2 & 44.1 & 43.9 \\
 35488 & 0.1$\pm$0.2 & $-$2.3$\pm$0.1 & 0.3$\pm$0.5 & 2.0$\pm$0.3 & 44.2 & 44.1 \\
 38108 & 0.5$\pm$0.1 & $-$1.9$\pm$0.1 & 0.8$\pm$0.2 & 3.3$\pm$0.2 & 44.7 & 44.7 \\
 41225 & 0.3$\pm$0.1 & $-$1.7$\pm$0.1 & 1.0$\pm$0.2 & 3.9$\pm$0.1 & 45.0 & 44.7 \\
 45924 & 0.5$\pm$0.1 & 0.0$\pm$0.1 & 1.6$\pm$0.3 & 1.1$\pm$0.9 & 44.9 & 45.3 \\
\enddata
\tablecomments{Table of continuum measurements from the PRISM spectroscopy. $^a$ is the highest S/N spectrum of the triply-imaged object from \citet{Furtak:2023nature}. The $A_V$ is derived from the Balmer decrement for this source. $^b$ is the $z=8.5$ broad-line AGN discussed by \citet{Kokorev:2023}; in this case the $A_V$ measurements are derived from the Balmer decrement and a full spectral fit respectively, while $L_{5100}$ in Col. (7) is derived from H$\beta$ assuming a ratio of H$\alpha$/H$\beta = 3.5$. 
Column (1): MSA ID. 
Column (2): Optical slope $f_{\lambda} \propto \lambda^{\alpha}$, fitted redward of 4400\AA.
Column (3): UV slope $f_{\lambda} \propto \lambda^{\beta}$, fitted blueward of 3000\AA. 
Column (4): $A_{\rm V}$ (mag) estimated from the continuum slope assuming an intrinsic AGN power-law slope of $\alpha_{\rm opt} = 0.45$ and an SMC reddening law. 
Column (5): $A_{\rm V}$ (mag) estimated from the continuum slope assuming an intrinsic AGN power-law slope as fitted to the UV component of the spectrum and an SMC reddening law. 
Column (6): Demagnified and dereddened luminosity at 5100\AA\ (erg/s) as estimated from the measured continuum and the $A_{\rm V}$ from Column (5).
Column (7): Demagnified and dereddened luminosity at 5100\AA\ (erg/s) as estimated from the measured H$\alpha$ luminosity and the $A_{\rm V}$ from Column (5).
}
\end{deluxetable*}

\begin{deluxetable*}{lllllllllllll}
\tabletypesize{\footnotesize}
\tablecolumns{13}
\tablewidth{0pt}
\tablecaption{H$\alpha$ fits\label{table:thefits}}
\tablehead{
\colhead{MSAID} & \colhead{FWHM$_n$} & \colhead{FWHM$_{\rm two}$} & \colhead{$f_{\rm two}$} & \colhead{$\chi^2_n$} & \colhead{$\chi^2_{\rm two}$} & \colhead{$L_{\rm H\alpha}$} & \colhead{$M_{\rm BH}$} & \colhead{$L_{\rm bol}$}\\ 
\colhead{(1)} & \colhead{(2)} & \colhead{(3)} & \colhead{(4)} & \colhead{(5)} & \colhead{(6)} & \colhead{(7)} & \colhead{(8)} & \colhead{(9)}
}
\startdata
 2008 & 650$\pm$100 & 1200$\pm$430 & 1.4$\pm$0.3 & 94 &  94 & \nodata & \nodata & \nodata & \\
 4286 & 800$\pm$10 & 2900$\pm$1040 & 13.7$\pm$0.4 & 807 &  361 & 43.4 & 8.0$\pm$0.3 & 45.4$\pm$0.3 & \\
 10686 & 60$\pm$10 & 4900$\pm$1570 & 5.5$\pm$1.6 & 332 &  296 & \nodata & \nodata & \nodata & \\
 13123$^a$ & \nodata & 2670$\pm$170 & 5.0$\pm$0.45 & \nodata & \nodata & 42.7 & 7.3$\pm$0.2 & 45.0$\pm$0.1 \\
 13821 & 790$\pm$10 & 3100$\pm$710 & 17.4$\pm$0.8 & 461 &  248 & 43.3 & 8.1$\pm$0.2 & 45.4$\pm$0.2 & \\
 20466$^b$ & 203$\pm$154 & 3439$\pm$413 & 2.3$\pm$0.2 & 4817 & 410 & 43.8 & 8.17$\pm$0.42 & 45.8$^{+0.3}_{-0.1}$ \\
 23608 & 70$\pm$10 & 2900$\pm$340 & 2.4$\pm$0.2 & 409 &  341 & 42.3 & 7.5$\pm$0.2 & 44.2$\pm$0.4 & \\
 28876 & 650$\pm$70 & 1800$\pm$650 & 2.2$\pm$0.4 & 265 &  265 & \nodata & \nodata & \nodata & \\
 35488 & 90$\pm$10 & 1900$\pm$210 & 10.9$\pm$2.1 & 1123 &  976 & 42.8 & 7.4$\pm$0.2 & 44.8$\pm$0.4 & \\
 38108 & 800$\pm$10 & 4100$\pm$1980 & 14.3$\pm$0.5 & 671 &  215 & 43.4 & 8.4$\pm$0.5 & 45.3$\pm$0.5 & \\
 41225 & 510$\pm$30 & 2000$\pm$600 & 4.7$\pm$0.3 & 539 &  435 & 43.5 & 7.7$\pm$0.4 & 45.3$\pm$0.5 & \\
45924 & 70$\pm$10 & 4500$\pm$40 & 383.4$\pm$0.4 & 1621505 &  90681 & 44.0 & 8.9$\pm$0.1 & 46.4$\pm$0.2 & \\
\enddata
\tablecomments{Table of H$\alpha$ measurement from the PRISM spectroscopy. $a$ is the highest S/N spectrum of the triply-imaged object from \citet{Furtak:2023nature}, while $b$ is the $z=8.5$ broad-line AGN from \citet{Kokorev:2023}.
Column (1): MSA ID. 
Column (2): FWHM~(H$\alpha$), narrow-only fit (km/s).
Column (3): FWHM~(H$\alpha$), two-component broad fit (km/s). (b) is measured from H$\beta$ \citep{Kokorev:2023}.
Column (4): Observed H$\alpha$ flux, two-component broad fit ($10^{-18}$ erg/s/cm$^2$). (b) is measured from H$\beta$ \citep{Kokorev:2023}. 
Column (5): $\chi^2$ (narrow-only). 
Column (6): $\chi^2$ (two-component). 
Column (7): De-magnified and dereddened H$\alpha$ luminosity (erg/s), using $A_{V, fit}$ from Table \ref{table:continuum}. Only definite broad-line objects have H$\alpha$ luminosities listed.
Column (8): Black hole mass, using H$\alpha$ linewidth and luminosity. Errors account linewidth errors and two reddening values. Maybe sources have no \mbh\ listed.
Column (9): Log of the bolometric luminosity (erg/s), estimated from the H$\alpha$ luminosity. Errors account for two reddening values and uncertainty in the bolometric correction.
}
\end{deluxetable*}

\subsection{Identifying Broad Balmer Lines}

Two targets (A2744-QSO1 and MSAID20466) have previously been identified as robust broad-line AGN \citep{Furtak:2023nature,Kokorev:2023}. We here examine the remaining 10 objects, after removing the three brown dwarfs. The majority of our targets have $z<7$, meaning that H$\alpha$ falls fully in the spectrum. Given the significant dust reddening, we focus exclusively on identifying broad H$\alpha$, since given the typical Balmer decrements of $\sim 5-10$, we have substantially higher signal-to-noise in H$\alpha$ than H$\beta$.

To determine which sources have broad lines, we apply the following three criteria. First, we insist that the fit improvement in $\chi^2$ be better than $3 \sigma$ ($\Delta \chi^2 > 11.5$ for four additional degrees of freedom) between the narrow-only and two-component fit (MSAID2008 and MSAID35488 are removed at this step). Second, we require that the two-component broad H$\alpha$ have FWHM$>2000$~km/s. We choose 2000~km/s as a conservative limit, compared to the bimodality at 1200 km/s identified by \citet{Hao:2005}. Finally, we remove any object with $<5 \sigma$ detection of a broad line, where $\sigma$ is measured as the 68\% distribution from the nested sampling (MSAID10686 is removed at this step). All objects have $>5 \sigma$ narrow \oiii\ and H$\alpha$ detections. The ``unconfirmed black hole'' targets could show evidence of broad emission lines with higher spectral resolution (and in one case complete coverage of H$\alpha$). Specifically, many very significant but weak broad lines have been identified in higher-resolution \jwst\ spectroscopy \citep[e.g.,][]{Larson:2023,Harikane:2023,Maiolino:2023}. We could not identify such lines in our data, and so we retain a label of unconfirmed for the remaining three objects. We do not report broad-line fluxes or black-hole masses for the unconfirmed targets.

Another question is whether the broad lines might arise from outflows, rather than the broad-line region of the AGN. It is true that outflows of $\sim 2000$~km/s have been seen in narrow permitted and forbidden lines in very luminous (and reddened) AGN at $z \sim 2$ \citep[e.g.,][]{Zakamska:2016}. However, while we do not have robust measurements of the narrow-line widths, we know that they are $<500$~km/s, so that the broad permitted lines almost certainly arise from the broad-line region.

\subsection{Other Emission Lines}

We do not present exhaustive fits of other UV emission lines in this work. However, we did perform a search for the forbidden line \nev$\lambda 3426$. This line is widely accepted as a likely indicator of AGN activity, given the ionization potential of 95 eV, although it has been attributed to Wolf-Rayet stars and/or shocks in various situations \citep[e.g.,][]{ 
Abel:2008,Izotov:2012,Leung:2021}. We only find compelling evidence for \nev\ in source 45924, which is already a very strong AGN candidate by virtue of the broad-line width of FWHM(H$\alpha$)$ = 4000$~km/s. Since \nev\ is typically $0.02-0.2$ the strength of \oiii, and considering the strong reddening, this non-detection is not surprising \citep{Netzer:1990}. We also attempt to decompose the \oiii$~\lambda 4363$ and H$\gamma$ emission lines, which may also provide corroborating evidence for excitation by an AGN \citep{Baskin:2005,Binette:2022}. However, given our spectral resolution, we do not achieve robust decompositions for any of the unconfirmed AGN candidates.

\subsection{Balmer decrement and continuum slopes}
\label{sec:reddening}

By selection, our targets contain steep red continua. With the spectra, we have good constraints on both the UV and optical slopes, which we will denote as $\beta_{\rm UV}, f_{\lambda} \propto{\lambda^{\beta}}$ ($\lambda < 3000$~\AA) and $\alpha_{\rm opt}, f_{\lambda} \propto{\lambda^{\alpha}}$ ($\lambda > 4000$~\AA). For the definitive and unconfirmed broad-line sources, we find $\langle \beta_{\rm UV} \rangle = -1.5 \pm 0.7$ and $\langle \beta_{\rm UV} \rangle = -1.9 \pm 0.2$ respectively. The unconfirmed AGN candidates are bluer in the rest-UV than the average AGN in our sample, but all the targets are on the redder end of what is seen for F444W selected galaxies with $4<z<7$ \citep{Bouwens:2016,Bhatawdekar:2021,Topping:2023,Nanayakkara:2023}. 
The optical slopes are by construction quite red, with $\langle \alpha_{\rm opt} \rangle = 0.5 \pm 0.3$, $\langle \alpha_{\rm opt} \rangle = 0.9 \pm 0.4$ for AGN and unconfirmed samples respectively (Table \ref{table:continuum}).

We now try to estimate the dust reddening. Nominally, we measure the Balmer decrement from the narrow Balmer line fits (ranging from 4-15), implying $A_V \sim 1-5$. Because of the degeneracy with the broad Balmer lines, the Balmer decrements are not well constrained in general. Small changes in the broad-line fits can lead to large fluctuations in the Balmer decrements. Note that we do not trust broad-line Balmer decrements, since self-absorption in the broad-line region can also change the H$\alpha$ to H$\beta$ ratio \citep{Korista:2004}. We thus base our reddening estimates on the continuum slopes. 

We fit the UV and optical sides of the spectra separately to derive $\beta_{\rm UV}$ and $\alpha_{\rm opt}$ as power-laws in $f_{\lambda}$. To derive the dust reddening, we must assume an intrinsic (unreddened) spectral shape. All of our determinations of $A_V$ are under the assumption that the red continuum is dominated by AGN light (see \S \ref{sec:lrd}). We take two AGN models. First, we use the Sloan Digital Sky Survey \citep[SDSS][]{York:2000} composite AGN templates from \citet{VandenBerk:2001}. The composite slope has $\beta_{\rm UV} = -1.5$ in the UV. The UV slope of the SDSS template is consistent with many other works \citep[e.g.,][]{Davis:2007}, and seems to hold over a wide range of redshift \citep{Temple:2021}. The SDSS template then has a break to $\alpha_{\rm opt} = 0.45$ in the rest-frame optical at $\lambda \gtrsim 4000$\AA. The redder slope could be an intrinsic spectral change, or the impact of galaxy light redward of the 4000\AA\ break, which is hard to determine robustly. Hence, we fit the SDSS-QSO template independently to each spectrum in the range $\lambda < 3500$\AA\ and $\lambda > 4300$\AA\ (to avoid the observed break seen in several of the spectra). In each case we allow a free extinction parameter characterized by the SMC reddening curve \citep{Gordon:2003}. 

We also perform a second more agnostic fit in which we fit the rest-frame UV slope directly assuming $F_{\lambda} \propto \lambda^{\beta}$, and then assume this empirical slope holds over the full spectral range. In the case that the UV emission arises from scattered light, a single spectral slope should describe the scattered and intrinsic slopes. We allow for an additional two free parameters, an $A_{\rm V}$ affecting an SMC-like reddened $F_{\lambda}$ component to describe the red end of the spectra, and a zero-reddening scattered light fraction ($f_{\rm scatt}$) that describes the rest-frame UV portion of the final model such that $f_{\lambda{\rm, model}} = f_{\rm scatt}F_{\lambda{\rm ,UV}}$.

The reddening derived from the SDSS template is tabulated as $A_{\rm V, 0.45}$ due to the rest-frame optical slope of the composite spectrum. The agnostic empirical fit is denoted as $A_{\rm V, fit}$ because we fit directly for $\beta$. These two reddening values thus bracket a reasonable range of reddening levels, and are presented in Table \ref{table:continuum}. 

We find that although the flux level in the UV is suppressed compared to the standard AGN template, the UV slopes are very similar to the unreddened UV slope of the SDSS AGN template ($\langle \beta_{\rm UV} \rangle = -1.5 \pm 0.7$). The rest-optical slopes are more variable. In \S \ref{sec:uvorigin}, we will discuss the origin of the UV emission.

\section{High fraction of AGN in red photometric samples}
\label{sec:yield}

Our NIRSpec/PRISM spectra confirm the AGN hypothesis for a majority of the objects photometrically selected in L23. With the spectra in hand, we can explore the demographics of red sources at high redshift, as well as devise refined photometric selection criteria for future searches.

\begin{figure*}
\hspace{2mm}
\includegraphics[width=0.9\textwidth]{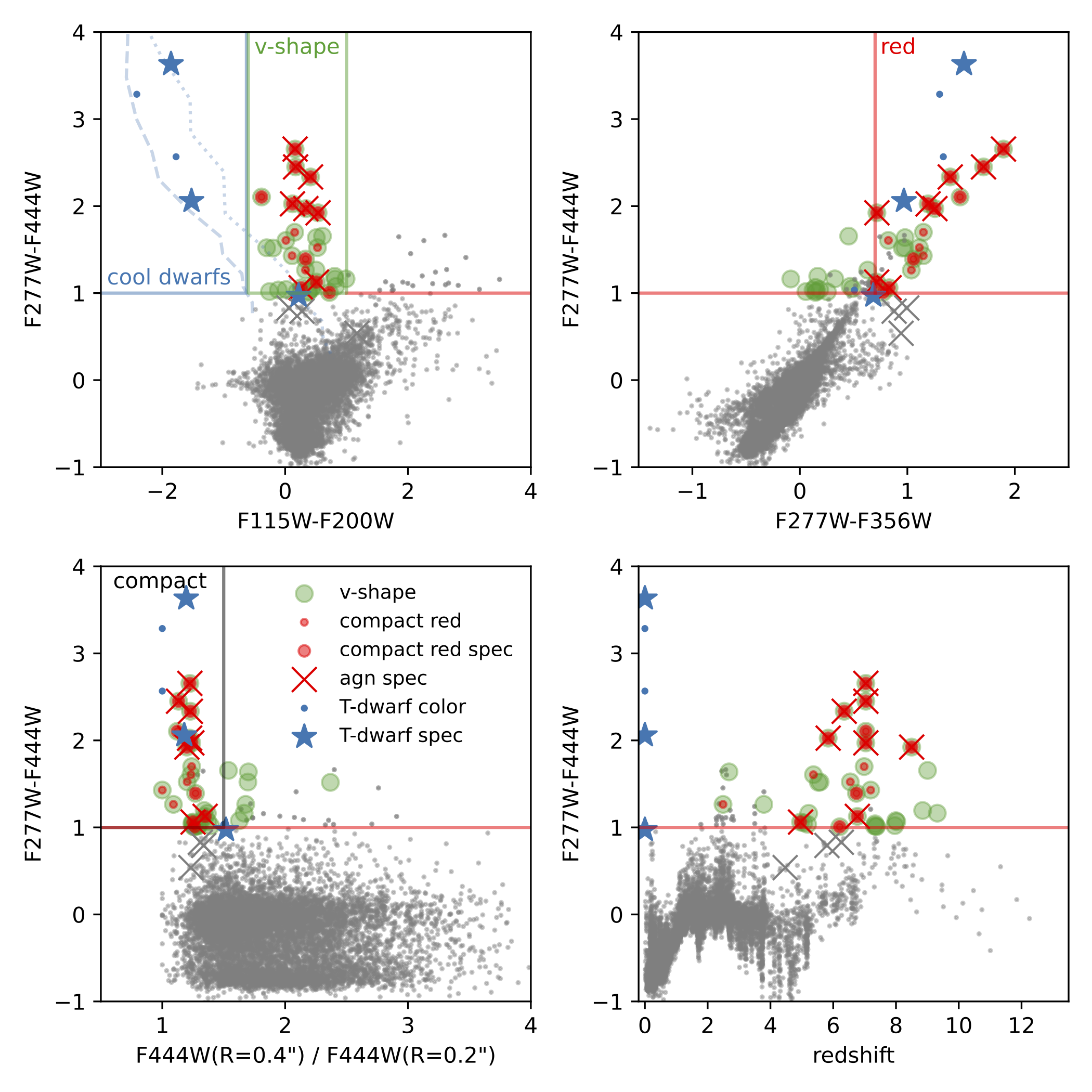}
\caption{AGN in the context of high-redshift red galaxies. The grey points are the UNCOVER catalog for F444W$>27.7$~mag and F444W S/N$>14$. Top left: NIRCam F115W$-$F200W versus F277W$-$F444W bi-color selection identifies  galaxies like those in \citet{Labbe:2023} that have blue rest-frame UV continuum and red rest-frame optical continuum (``v-shape'' SED). The AGN candidate sample of \citet{Labbe:2023uncover} is a subset of that ($\sim50\%$), selected to also have red optical sloped continuum via a cut in two adjacent filters (top right, and compact size (bottom left), using a ratio of aperture fluxes as a proxy for size. Brown dwarf star contaminants generally have bluer F115W$-$F200W than galaxies and can therefore be isolated. Overplot are synthetic color tracks from the LOWZ brown dwarf atmosphere models \citep{2021ApJ...915..120M} for T$_{\rm eff}$ $\leq$ 1600~K and solar [M/H] = 0 and $-1.5$. The $v-shape$ criterion alone is very effective at selecting for $z>5$ galaxies. A selection including the compact + red criterion is efficient at selecting red AGN.
}
\label{fig:newsample}
\end{figure*}

\subsection{Yield and Contaminants}

Of the 17 targets from L23 with UNCOVER PRISM spectra, 11 have unambiguous evidence for broad emission lines, including three which are multiple images of the same target \citep{Furtak:2023nature}. Three targets do not show clear evidence for broad emission lines, and a further three are brown dwarfs \citep{Burgasser:2023}. Therefore 11/17 $= 65\%$ of the sample are confirmed as broad line AGN. Accounting for multiplicity, the confirmed AGN fraction among the targetted extragalactic objects in our sample is 9/12 $= 75\%$, but we note AGN activity is not strongly ruled out in the remaining galaxies, so the true fraction could be as high as $100\%$.

One obvious contaminant are brown dwarf stars. A simple color cut, excluding all sources bluer than F115W$-$F200W$< -0.5$ (blue box in Figure \ref{fig:newsample}), would remove the majority of brown dwarfs from high-z galaxy selections in high-latitude deep fields \citep[see also][]{Langeroodi:2023}. Based on the simulations from \citet{Burgasser:2023} in the UNCOVER area we expect $\sim$ 5 low-metallicity halo brown dwarfs, expected to have the bluest F150W$-$F200W colors, as can be seen from model tracks in Figure \ref{fig:newsample}. Some contaminants at redder colors would remain: $\sim 2$ metal-rich brown dwarfs that reside in the Galactic thin disk are expected. The higher metallicity brown dwarfs in the thin disk should be bright, F444 $< 24$~mag, since the limited vertical scale height of the disk will limit these brown dwarfs to be within a few hundred pc. A more refined filter could possibly be defined from a larger model and/or template sample. Upcoming medium-band data (GO-4111; PI W. Suess) will allow for even cleaner identification of brown dwarfs \citep{Hainline:2023bd}.

\subsection{Updated Photometric Selections of High Redshift Galaxies and AGN}

Here we define updated NIRCam-only criteria to the selections of \citet[][]{Labbe:2023uncover,Labbe:2023}. Specifically, we start with the same $SNR(F444W) > 14 \ \& \ m_{F444W} < 27.7$~mag cuts as before. Then, we define a ``v-shape'' color selection to mimic that of \citet[][]{Labbe:2023}, designed to find candidate massive high-redshift galaxies:

$${\it v-shape}  $$ 
$$ (-0.5 < \rm{F115W-F200W} < 1.0) \ \land $$
$$ (\rm{F277W-F444W} > 1.0)  $$

The main difference with respect to \citet[][]{Labbe:2023} is using $\rm{F115W-F200W}$ rather than $\rm{F150W-F200W}$, and a blue limit to facilitate removing brown dwarfs. In addition, since we lack deep ACS coverage, we forego the \emph{HST}/ACS optical nondetection criterion, thereby extending the selection towards lower redshift $z\sim5$. This selection produces 31 unique sources in UNCOVER and is both effective and complete at identifying high-redshift galaxies with red rest-frame optical colors. The median $<z>=6.9 \pm 1.0$ and 28/31 = $90\%$ sources have $z_{phot}>5$ (see Figure \ref{fig:newsample}). In total 15 sources were targeted with UNCOVER PRISM spectroscopy, 11 presented in this paper and a further 4 presented in Price et al. (in preparation), all are at $5 < z < 8.5$. Only one source with F277W$-$F444W$>1$ at $z>5$ does not satisfy the v-shape selection.

Remarkably, with this color selection alone, at least 29\% of the objects are spectroscopically identified AGN, corresponding to about one-third of all F277W$-$F444W$>1$ at $z>5$. We expect there are many non-active galaxies in the sample as well. Most of the \citet{Labbe:2023} objects found in the CEERS field are spatially resolved in the rest-UV, with densities similar to the cores of present-day ellipticals \citep{Baggen:2023}. Out of 4 galaxies with spectroscopy in that sample, one was found to be a red broad line AGN \citep{Kocevski:2023}. Cycle 2 spectroscopy (program JWST-GO-4106, PI Nelson) will determine what fraction of these color-selected but spatially-resolved objects show evidence for AGN activity. 

To select red AGN more specifically in a similar fashion to the $red2$ selection of L23, we add an additional red color and compactness criterion compared to the v-shape criterion:

$${\it compact \ red}  $$ 
$$ (-0.5 < \rm{F115W-F200W} < 1.0) \ \land $$
$$ (\rm{F277W-F444W} > 1.0)   \ \land $$
$$ (\rm{F277W-F356W} > 0.7) \,   \ \land  $$
$$ {\it compact} = f_{\rm{F444W}}(0.4\arcsec) / f_{\rm{F444W}}(0.2\arcsec) < 1.5 $$

where the additional $\rm{F277W-F356W}>$ 0.7 facilitates selecting SEDs with red continuum slopes, rather than breaks or emission lines (Figure \ref{fig:newsample}, top right), and the compactness criterion helps to target point-source dominated sources (Figure \ref{fig:newsample}, bottom left). These extra cuts remove roughly 50\% of the $v-shape$ targets. The main differences compared to L23 are the $\rm{F115W-F200W} > -0.5$ brown dwarf removal and a slightly more stringent $compact$ cut than in L23.

\begin{figure}
\hspace{0.1mm}
\includegraphics[width=0.4\textwidth]{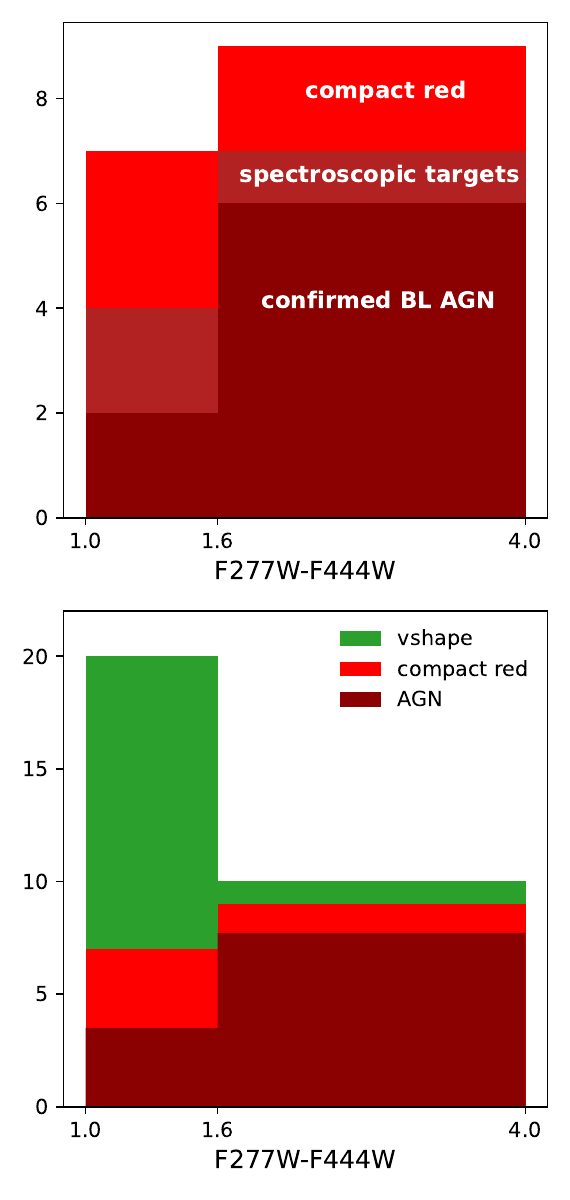}
\caption{Distribution of object types as a function of F277W$-$F444W color. Top panel shows the compact+red NIRCam selected sample, the number of spectroscopic targets, and the confirmed BL AGN. Bottom panel shows targets selected by "v-shape" two-color cut (green), of which $90\%$ are at $z=5-10$, the number of compact+red selected sources (red), and estimated number of AGN (dark red).}
\label{fig:agnfraction}
\end{figure}

Amongst this more stringent AGN-focused selection, we find at least 75\% are AGN. The AGN fraction is a function of F277W$-$F444W color (Figure \ref{fig:agnfraction}). The majority of the broader sample of $v-shape$ galaxies have $1 < \rm{F277W-F444W} < 1.6$. Therefore, an alternative high yield selection of AGN can be made by selecting on F277W$-$F444W$>1.6$. In this situation, $>$80\% of the galaxies are AGN, while $>90$\% of the compact red galaxies are AGN \citep[see also ][]{Barro:2023}.

In short, among $z>5$ galaxies with red colors ($-0.5<$F115W--F200W$<1.0$, F277W--F444W$>1$), at least one third are AGN. Making more stringent compactness and color criteria, or simply cutting at F277W--F444W$>1.6$, can yield an 80-90\% AGN fraction.

\section{The Nature of the `Little Red Dots'}
\label{sec:lrd}

Now focusing exclusively on the confirmed broad-line AGN, we explore their properties, including SEDs, luminosity functions, and black holes masses.

\subsection{Origin of the Continuum}
\label{sec:uvorigin}

\begin{figure*}
\hspace{+5mm}
\includegraphics[width=0.8\textwidth]{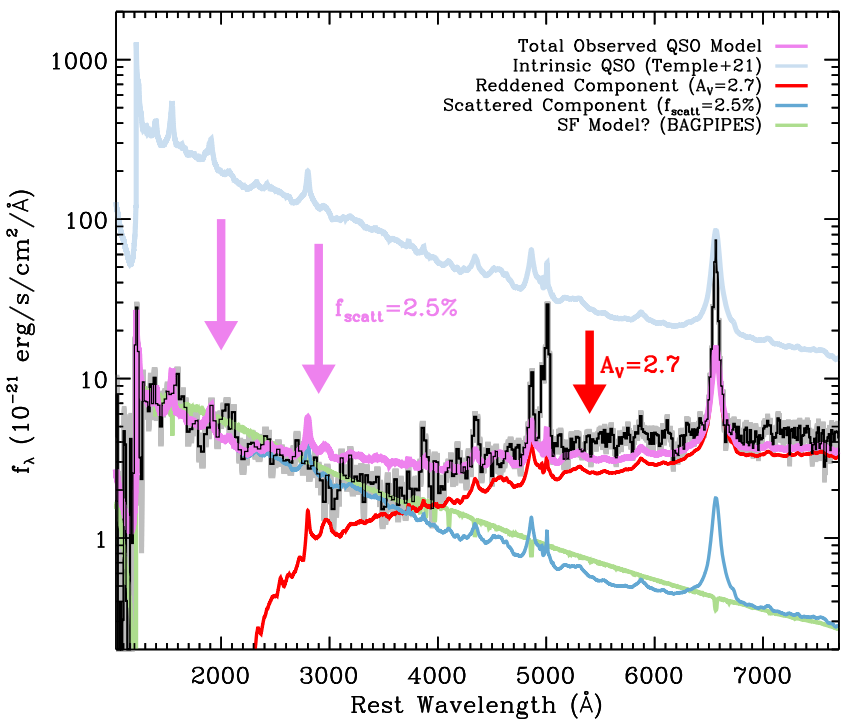}
\caption{We illustrate our preferred model for the particular red and UV slopes seen in our objects using MSAID4286. The intrinsic AGN continuum (red) is highly reddened. Thus, the UV component cannot be explained by the primary AGN continuum. Here, we explore the possibility that the UV comes from scattered light, at 2.5\% of the intrinsic UV (shown schematically in light blue). For illustration, we show here the \citet{Temple:2021} template. However, we achieve even better fits when we use the observed UV slope as the intrinsic power-law AGN shape. We also overplot a stellar-population fit to the UV-side of the spectrum with \texttt{Bagpipes} \citep{Carnall:2018, Carnall:2019}, again to illustrate that with moderate star-formation rates of a few solar masses per year, and $A_V \sim 0.6$~mag, it is possible to fit the UV continuum slope with star light as well.
}
\label{fig:uvfit}
\end{figure*}

The compact red objects were photometrically selected not only to have a red continuum in the rest-frame optical, but also detectable UV emission. From a template fit to the photometry alone, we came to the conclusion that the red continuum was most likely dominated by reddened AGN light. With the spectra, we now ask whether it is possible to distinguish the origin of the rest-frame optical and UV continua more directly. We demonstrate the different methods of modeling the spectra with one example in Figure \ref{fig:uvfit}. We will defer full SED modeling to a future work, as that will require exploiting both the continua and the emission lines.

In the rest-frame optical, we immediately see that the observed red slopes are continuum-dominated, not arising from high-EW emission lines \citep[see possibilities in][]{Furtak:2022,Endsley:2023}. The rest-frame optical slopes are consistent with a reddened AGN template or dusty star-formation, as was concluded from the photometric fitting. However, the broad emission lines provide an additional clue. We can calculate the expected continuum given the observed H$\alpha$ luminosity \citep{GreeneHo:2005} and a measurement of reddening. We can also directly calculate the observed continuum. The two should match.\footnote{Comparing to the \citet{GreeneHo:2005} relation is a bit easier than looking at the H$\alpha$ equivalent width distribution directly, since there may be subtle trends between line equivalent widths and luminosity \citep[e.g.,][]{Croom:2002,Stern:2012a} that are fitted for directly by Greene \& Ho.} We include the two $L_{5100}$ values, based on the continuum and H$\alpha$ but using the same $A_{\rm V, fit}$, in Table \ref{table:continuum}. While the H$\alpha$-derived values tend to be a bit higher, they agree within a factor of two, suggesting that the broad H$\alpha$ equivalent widths are comparable to the low-redshift SDSS calibration sample from \citet{GreeneHo:2005}. The equivalent widths from pure star formation are much higher for dusty massive galaxies at lower redshift \citep{Fumagalli:2012,Whitaker:2014}. Given the detection of broad H$\alpha$ and the AGN-like ratios of H$\alpha$ to continuum, we conclude that the rest-frame optical is AGN dominated. 

The dominant component from the AGN is the red continuum, and given the $A_V = 1-3$ that we infer, we could not detect that component in the UV (see Fig.\ \ref{fig:uvfit}). Instead, we must invoke a second component, which is brighter than the pure reddened AGN spectrum, but still only a few percent of an unreddened (blue) AGN. The two possible origins for this second UV-emitting component are either star formation from the host or some small fraction of the photons from the accretion disk, whether via scattering or direct leakage. Unfortunately, we cannot make a concrete determination of the origin of the UV based on the spectral slopes alone. The rest-frame UV slopes are quite consistent with the observed slopes of blue AGN with $A_V \approx 0$, and in some cases there are likely broad UV lines. With higher resolution data this would be a smoking gun of AGN light dominating the UV. Alternatively, they could be consistent with the redder end of $4<z<7$ galaxies \citep[e.g.,][]{Bouwens:2016,Bhatawdekar:2021}, particularly those selected at F444W \citep[e.g.,][]{Nanayakkara:2023,Topping:2023}.

To explore the UV origin further, we can ask what UV luminosities we would expect given the intrinsic $L_{5100}$ luminosities. We bracket the range of reasonable $A_V$ using the two assumed intrinsic AGN slopes, and over that range we find that the observed $L_{3000}$ is 1-3\% percent or 10-20\% of percent of the expected intrinsic $L_{3000}$, depending on whether we assume $\alpha_{\rm opt, intrinsic} = \beta_{\rm UV, measured}$ or $\alpha_{\rm opt, intrinsic} = 0.45$ respectively. The former can easily be explained by a scattered (or directly transmitted) AGN component, while 10-20\% fractions are likely too high for pure AGN scattering \citep[e.g.,][]{Liu:2009}. In such cases, we would favor a star-formation contribution to the UV.  We choose not to fit the UV alone, since it has limited sensitivity to the stellar mass, but naively converting the $M_{1450}$ measurements to star-formation rates, and assuming that all the UV arises from star-formation, we find rates of $\langle {\rm SFR} \rangle = 1.5$~\msun/yr. In terms of $L_{\rm H \alpha}$, we would expect 10-1000 times less H$\alpha$ luminosity for this star formation rate than what we measure in the broad lines. On the other hand, this level of star formation seems both plausible and challenging to rule out from other measurements. In principle, the UV emission lines could distinguish the UV origins; higher resolution data are needed for this purpose.

Overall, we conclude that the rest-frame optical continuum is dominated by a dusty broad-line AGN continuum, and that more work is needed to definitively determine the origin of the UV emission. Finally, we also note that two objects, MSAID45924 and the triply-imaged A2744-QSO1 (MSAID13123 in this work) both have rather extraordinary spectral breaks that are not easily modeled with any AGN template. We leave to future work a more exhaustive fitting of the continuum, to attempt to understand the nature of their extreme breaks. 

\begin{figure*}
\hspace{0.1mm}
\includegraphics[width=0.48\textwidth]{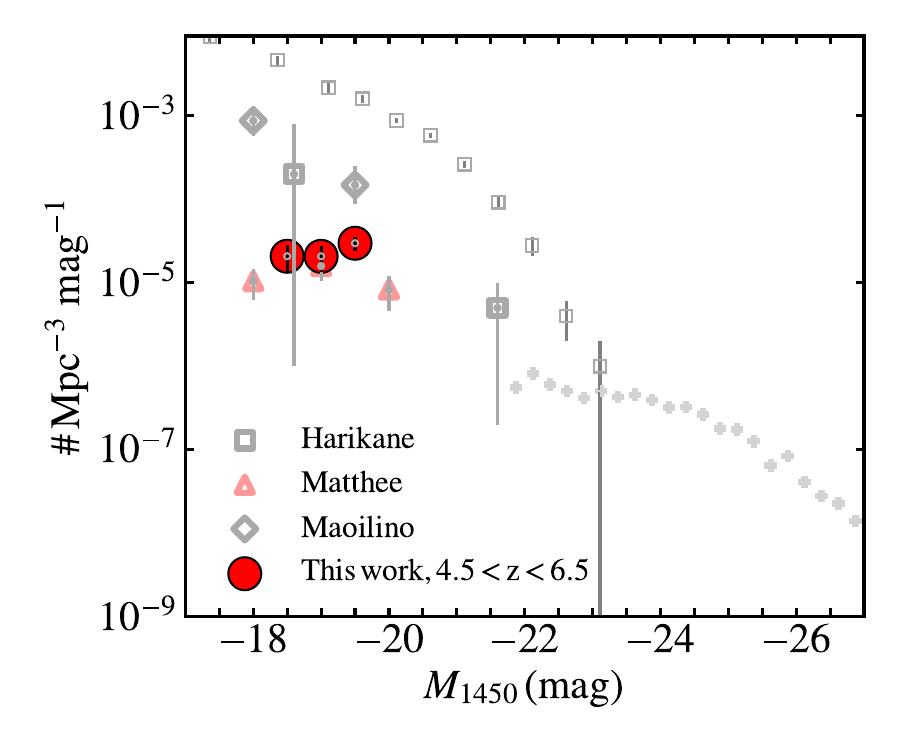}
\includegraphics[width=0.48\textwidth]{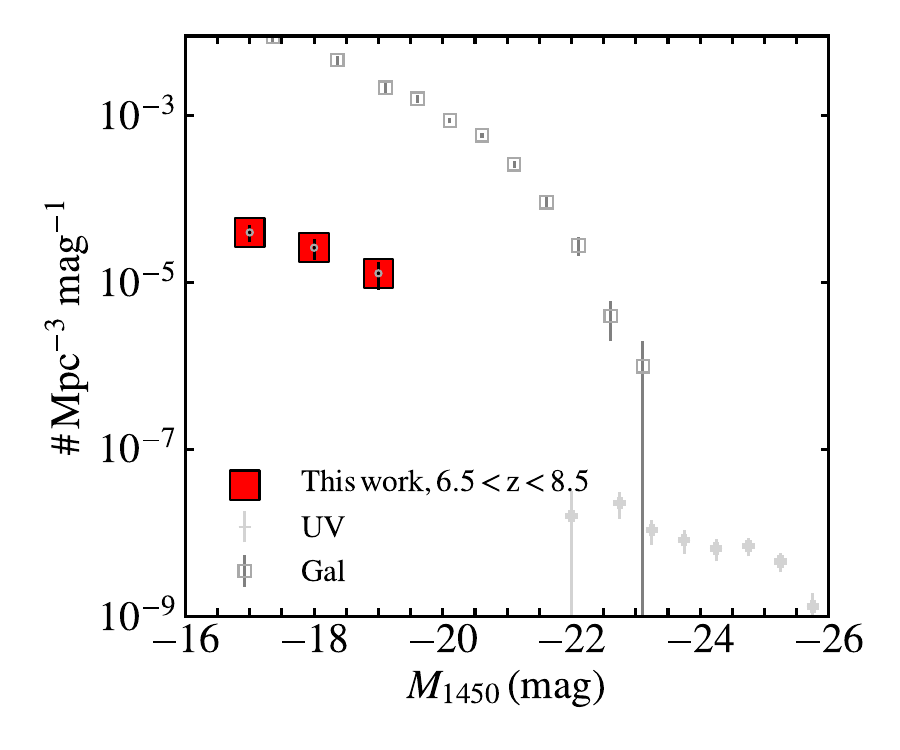}
\caption{UV luminosity function as measured at 1450\AA. We show the luminosity function in two redshift bins, $4.5<z<6.5$ in red circles and $6.5<z<8.5$ in red squares. We compare with the UV-selected luminosity functions from \citet{Akiyama:2018} (left) and \citep{Matsuoka:2023} (right). We show other \jwst-selected broad-line AGN from \citet{Harikane:2023}, \citet{Maiolino:2023}, and \citet{Matthee:2023}. Finally, we compare with the galaxy luminosity function from \citet{Bouwens:2017}. Consistent with Harikane et al., we find that the reddened AGN account for $\sim 20\%$ of the broad-line objects at this redshift, and a few percent of the galaxy population. Our AGN are far more numerous than the UV-selected ones, although they have overlapping bolometric luminosities. 
}
\label{fig:UVLF}
\end{figure*}

\subsection{Luminosity Functions}

\begin{table}
    \centering
    \caption{Rest-frame UV luminosity functions of our sample of red AGN. The UV luminosity bins have widths of 0.5\,magnitudes.}
    \begin{tabular}{ccc}
     $M_{\mathrm{UV}}$ &   $N$  &  $\phi(M_{\mathrm{UV}})~[\mathrm{Mpc}^{-3}\,\mathrm{mag}^{-1}]$\\\hline\hline
     \multicolumn{3}{c}{$z\sim5-6$ \textit{sample}}\\\hline
     $-$19.5                  &   $3$  &  $(3.0\pm0.6)\times10^{-5}$\\
     $-$19.0                  &   $2$  &  $(2.1\pm0.7)\times10^{-5}$\\
     $-$18.5                  &   $2$  &  $(2.1\pm0.8)\times10^{-5}$\\\hline
     \multicolumn{3}{c}{$z\sim7-8$ \textit{sample}}\\\hline
     $-$19.0                  &   $1$  &  $(1.3\pm0.5)\times10^{-5}$\\
     $-$18.0                  &   $2$  &  $(2.6\pm0.7)\times10^{-5}$\\
     $-$17.0                  &   $2$  &  $(4.0\pm1.0)\times10^{-5}$\\\hline
    \end{tabular}
    \label{tab:UV_LF}
\end{table}

\begin{table}
    \centering
    \caption{Bolometric luminosity functions of our sample of red AGN. The bins have widths of 1\,dex.}
    \begin{tabular}{ccc}
     $\log(L_{\mathrm{bol}}/\mathrm{erg}\,\mathrm{s}^{-1})$ &   $N$ &   $\phi(L_{\mathrm{bol}})~[\mathrm{Mpc}^{-3}\,\mathrm{dex}^{-1}]$\\\hline\hline
     \multicolumn{3}{c}{$z\sim5-6$ \textit{sample}}\\\hline
     $44.0$                                                 &   $1$ &   $(1.0\pm0.4)\times10^{-5}$\\
     $45.0$                                                 &   $4$ &   $(4.2\pm0.7)\times10^{-5}$\\
     $46.0$                                                 &   $1$ &   $(1.0\pm0.6)\times10^{-5}$\\\hline
     \multicolumn{3}{c}{$z\sim7-8$ \textit{sample}}\\\hline
     $45.0$                                                 &   $2$ &   $(2.6\pm0.5)\times10^{-5}$\\
     $46.0$                                                 &   $1$ &   $(1.3\pm0.5)\times10^{-5}$\\\hline
    \end{tabular}
    \label{tab:Lbol_LF}
\end{table}


\begin{figure*}
\hspace{-1mm}
\includegraphics[width=0.45\textwidth]{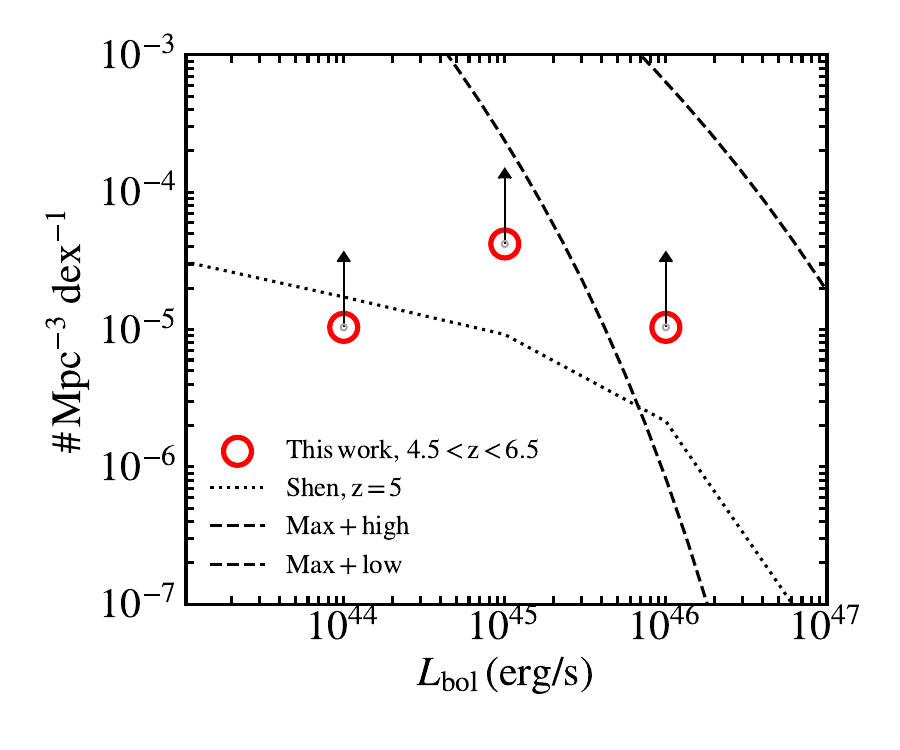}
\includegraphics[width=0.45\textwidth]{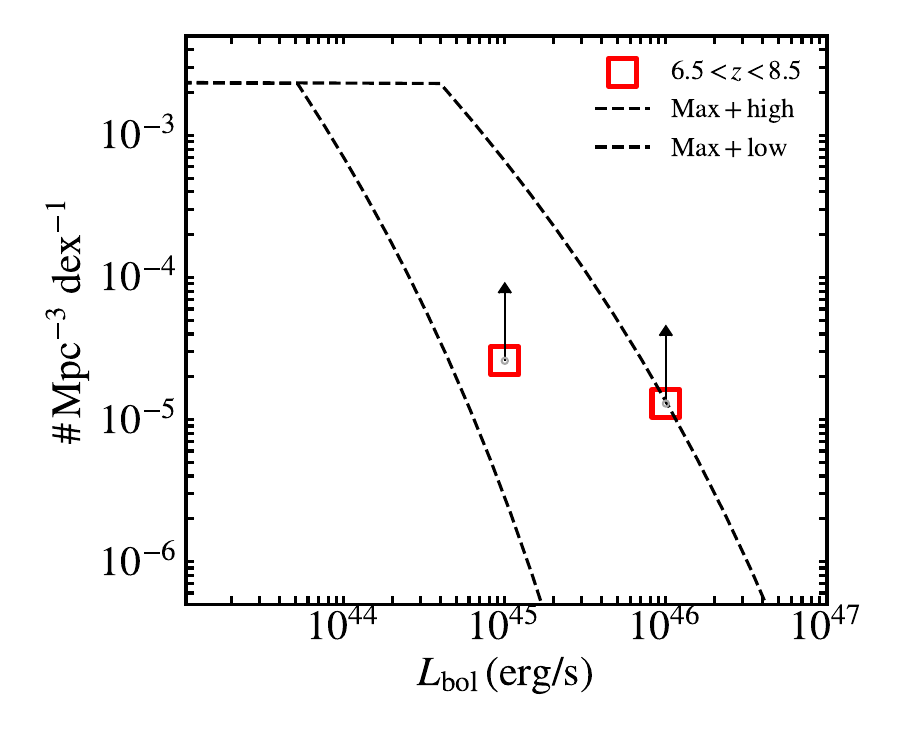}
\caption{Bolometric luminosity functions ($4.5<z<6.5$ left, $6.5 < z< 8.5$ right) as inferred from $L_{\rm H\alpha}$ (Table \ref{table:thefits}). The number densities are lower limits, particularly at low bolometric luminosity where our search will be particularly insensitive to galaxy-dominated objects. At left, we compare with the theoretical bolometric luminosity functions from \citet{Shen:2020}, using their ``local'' fits to $z=5$ data. In addition, we show a maximal bolometric luminosity function assuming that every halo harbors an accreting black hole radiating at its Eddington limit; the black hole mass is set from two local black hole scaling relations (see text for details). Under these assumptions, we are pushed towards an occupation fraction of unity at the highest $L_{\rm bol}$ that we probe, particularly in the higher redshift bin. 
}
\label{fig:LFbol}
\end{figure*}

We now compute the rest-frame UV luminosity function of red AGN at high-redshifts based on the spectroscopic sample alone. Since Abell~2744 is a strong lensing field, the lensing distortion needs to be taken into account when computing the volume of each luminosity bin. To calculate the volumes, we follow the forward-modeling method used in the \textit{Hubble Frontier Fields} by \citet{Atek:2018}: The sample completeness is assessed with a series of completeness simulations in which we populate the source plane with mock red AGN using the L23 SED, normalizing it to random UV luminosities and redshifts. These are then deflected into the lens plane with the deflection maps of the UNCOVER strong lensing model (see section~\ref{sec:SL}) and added into the mosaics on which we re-run the detection routines and assess the fraction of recovered sources to derive the selection function $f(z, M_{\mathrm{UV}})$. Note that the details of our completeness simulation methods will be published in Chemerynska et al. (in prep.). The selection function is used to weight the co-moving volume element which is then integrated over the source plane area of sufficient magnification required to detect each object in all bands given the UNCOVER mosaic depths (see equation~2 in \citealt{Atek:2018}) to obtain the effective volume probed by UNCOVER. Note that since the UNCOVER source detection is performed in the stacked LW bands \citep[see e.g.][]{Weaver:2023} where the compact red objects are the brightest, our sample is complete down to $M_{\mathrm{UV}}\sim-16$~mag and objects brighter than $M_{\mathrm{UV}}\sim-17$~mag do not necessarily need to be magnified in order to be detected given the UNCOVER depths listed in \citet{Weaver:2023}. Our sample is binned in UV luminosity bins of 0.5~magnitude width. The number count uncertainties are derived by drawing $10^4$ random luminosities from each object's $M_{\mathrm{UV}}$ error distribution and re-binning the luminosity function each time to allow objects to change luminosity bin. The uncertainty in magnification is taken into account in the computation of the UV luminosity uncertainties.

The resulting UV luminosity functions in the two redshift bins are presented in Table~\ref{tab:UV_LF} and Fig.~\ref{fig:UVLF}. As established above, the spectroscopic sample represents a conservative but fairly accurate proxy for the true luminosity function of the reddened broad-line AGN. We confirm the result from L23 that the number densities of these red-selected AGN are higher by roughly two orders of magnitude compared to the UV-selected AGN at similar magnitudes. The number density we find is also comparable to what has been inferred for other moderate-luminosity red AGN samples at $z \approx 5$ \citep[][]{Kocevski:2023,Barro:2023,Matthee:2023}, and accounts for $\sim 10-20\%$ of the general broad-line AGN population as selected with \jwst\ \citep{Harikane:2023,Maiolino:2023}. The spectroscopically-identified samples rely on higher-resolution NIRSpec data, which is inclusive of AGN with narrower lines and systems where the AGN does not necessarily dominate the total light output.

We have tried to emphasize that the UV light is a small fraction of the total luminosity due to the large reddening values and has an unknown origin in either scattered or transmitted AGN light or low-level unobscured star formation. Thus, while useful to put our targets into context, the UV luminosity function does not truly describe the physical properties of the AGN in this sample. For this reason, we also present a bolometric luminosity function in Table \ref{tab:Lbol_LF} and Figure \ref{fig:LFbol}. Since the derivation of completeness as a function of bolometric luminosity is not straightforward and would require detailed SED-modeling in the completeness simulations, we here use the fact that our sample is mostly complete in UV luminosity (see above) to approximate the effective volume for the bolometric luminosity function by assuming the maximum $M_{\mathrm{UV}}$-completeness derived above in each $L_{\mathrm{bol}}$-bin (of width 1\,dex) and no magnification, i.e. the effective volume element is integrated over the whole UNCOVER source plane area of $\sim27\,\mathrm{arcmin}^2$ \citep[see][]{Furtak:2022b_SLmodel,Atek:2023}. While this is a reasonable approximation given the properties of our sample, the thus derived bolometric luminosity function remains a lower limit. The lower luminosity bins in particular are probably underestimated since these would be more sensitive to magnification. 

To guide our interpretation, we also calculate ``maximal" black hole bolometric luminosity functions using a combination of high and low-redshift observations. We start by using the Sheth-Tormen halo mass function \citep{sheth-tormen2002} at $z \sim 4-7$. Each halo is assumed to contain a gas mass that is linked to the halo mass through the cosmological ratio such that $M_g = (\Omega_b/\Omega_m)M_h$. The total stellar mass in any halo is assumed to scale with the gas mass as $M_* = \epsilon_* M_g$. Matching to the evolving observed stellar mass function at $z \sim 4-7$ requires a star formation efficiency that evolves as $\epsilon_*= fn(z) = 0.15-0.03(z-6)$ at these redshifts \citep[for detailes see]{dayal2023}. For each halo, we then use two sets of observed relations linking the black hole and stellar mass at $z \sim 0$ from \citet{reinesvolonteri2015}: our ``high'' black-hole mass model uses $M_{BH} = 1.4M_* - 6.45$, valid for high stellar mass ellipticals. The ``low'' black-hole model on the other hand uses $M_{BH} = 1.05M_* - 4.1$, valid for moderate-luminosity AGN in low-mass halos. Based on previous theoretical works \citep{bower2017,dayal2019} that find suppressed Eddington accretion rate for black holes in halos below $10^{11}M_\sun$, we limit the Eddington accretion rate to $7.5\times 10^{-5}$ for such halos; higher mass halos are allowed to accrete at the Eddington rate in order to obtain the bolometric luminosity function shown in Fig. \ref{fig:LFbol}.

We can already draw some interesting conclusions from these bolometric luminosity function lower limits. First, we see the full range of $L_{\rm bol}$ spanned by the sample is quite broad, $10^{44} - 10^{47}$~erg/s, corresponding to Eddington-limited accretion onto black holes with \mbh$\sim 10^6-10^9$~\msun, most of the supermassive black hole population. We also see that the number density of sources at the bright end is approaching the maximal number of accreting black holes. In other words, we are finding an unexpectedly large number of relatively luminous black holes \citep[e.g.,][]{Furtak:2023nature,Kokorev:2023,Pacucci:2023}.

\subsection{Black Hole Masses}

We follow \citet{GreeneHo:2005}, as updated by \citet{reinesetal2013}, to calculate BH mass based on the luminosity $L_{\rm H\alpha}$ and velocity FWHM(H$\alpha$) of the broad H$\alpha$ line. These single-epoch BH mass estimates \citep[e.g.,][]{Shen:2019} are based on assuming that the BLR acts as a dynamical tracer of the BH \citep[e.g.,][]{Pancoast:2014}. The size of the BLR is estimated from the AGN luminosity \citep[e.g.,][]{bentzetal2013}, and then assuming virial equilibrium, the dynamical mass scales as \mbh$\propto \rm{FWHM_{H\alpha}^2 R}$. Of course, we do not know that the BLR is in virial equilibrium, nor do we know whether we probe the velocity field at a comparable radius as the BLR ``size'' we estimate from the luminosity \citep[e.g.,][]{krolik2001,Linzer:2022}. 

\begin{figure*}
\hspace{0.1mm}
\includegraphics[width=0.48\textwidth]{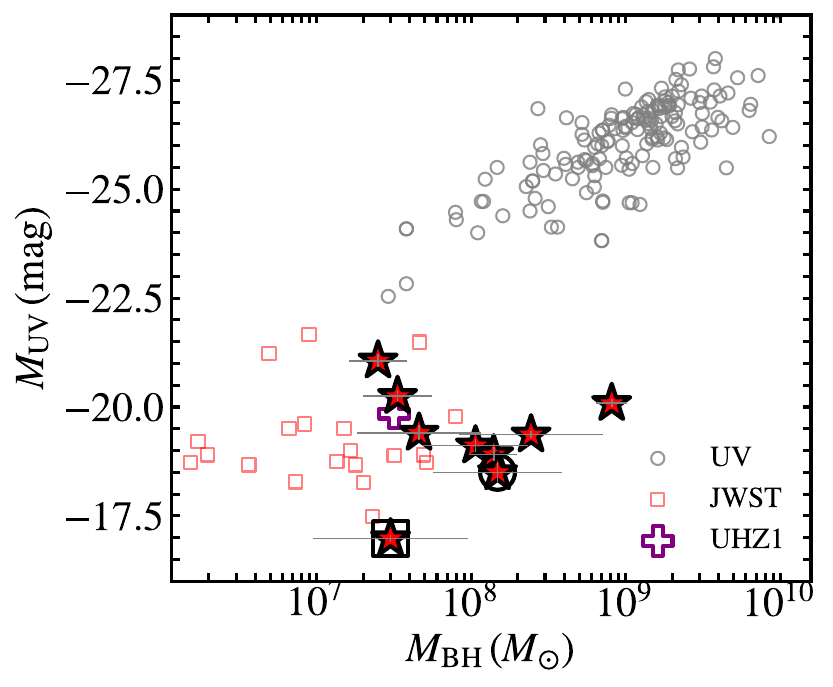}
\includegraphics[width=0.48\textwidth]{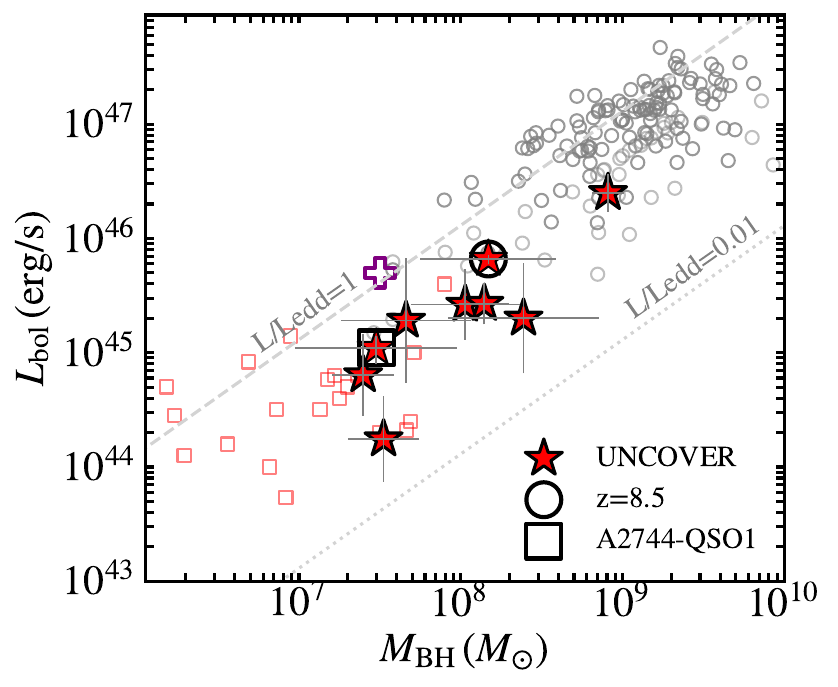}
\caption{Black hole mass versus $M_{\rm UV}$ (left) and bolometric luminosity (right) for the broad-line AGN (red stars) including A2744-QSO1 \citep[][; red star+square]{Furtak:2022,Furtak:2023nature} and the source MSAID20466 at $z=8.5$ \citep[][; red star+open circle]{Kokorev:2023}. For context, we include the UV-selected AGN with $z>5$ from \citet{Fan:2023}, other \jwst-selected broad-line sources \citep{Harikane:2023,Matthee:2023,Maiolino:2023}, and the X-ray--detected AGN at $z=10.XX$, UHZ1 \citep{Bogdan:2023,Goulding:2023}. Note that our sources extend to surprisingly high \mbh$=10^9$~\msun. 
}
\label{fig:mbhlbol}
\end{figure*}

The black hole masses are plotted against the UV luminosities and the H$\alpha$-inferred bolometric luminosities in Figure \ref{fig:mbhlbol}. We compare the sources both to luminous UV-selected quasars from the review of \citet{Fan:2023}  and from recent moderate-luminosity broad-line AGN discovered with \jwst\ \citep{Kocevski:2023,Harikane:2023, Barro:2023,Matthee:2023}. Our sources are on the massive end of the broad-line AGN found in deep \jwst\ fields, but barely reach the low end of the BH mass and luminosity seen in the rare UV-selected sources. However, their UV luminosities are 4-5 magnitudes lower than the UV-selected AGN at comparable \mbh. This difference is very likely due to dust obscuration. As shown on the right hand
side, when we use the broad emission lines to estimate the bolometric luminosity, we find much more agreement in luminosity range at a given \mbh. 

It is worth noting MSAID45924. This galaxy is the brightest in the sample (F444W$=22$~mag) and stands out for its high SNR and black hole mass of \mbh$\approx 10^9$~\msun. The object warrants bespoke analysis that is beyond the scope of this work.

\section{Discussion and Summary}
\label{sec:discussion}

In this work, we present NIRSpec/PRISM spectroscopic follow-up of 15 red, compact sources selected in the UNCOVER+Abell 2744 field. The majority of these targets are confirmed to be AGN with $z>4.5$. The UV/optical SEDs have a characteristic steep red continuum towards the rest-frame optical but also a non-negligible UV component. The rest-frame optical is consistent with a reddened ($A_V \sim 1.5$) broad-line AGN. The UV slopes are well fit as unobscured AGN in slope, but suppressed by $\sim 100$ times relative to an un-reddened source. From the available low-resolution spectroscopy and broad-band SED data, we cannot rule out that the UV component is due to moderately reddened star formation at the level of a few solar masses per year in the host. AGN with similar shapes \citep{Noboriguchi:2023} are known at all redshifts \citep[e.g.,][]{Veilleux:2016,Pan:2021,Glikman:2012,Banerji:2015,Hamann:2017,Assef:2018}. However, at $z<3$, these reddened sources with a UV excess are rare; \citet{Noboriguchi:2019} estimate that of all the dust-obscured galaxies, only 1\% have a UV excess. 

\jwst\ is uncovering a surprisingly high number density of red AGN at $z>5$ \citep[see also][L23]{Harikane:2023,Barro:2023,Matthee:2023}. This high number density is unexpected compared to UV-selected sources, which have measured number densities nearly 100 times lower at their faintest UV luminosities \citep[e.g.,][]{Matsuoka:2018,Matsuoka:2023}, although nominally the densities are similar to some X-ray selections \citep{Giallongo:2019}. Additionally, the compact red AGN account for a large fraction of all red-selected sources with \jwst. After applying simple color cuts designed to select massive galaxies \citep[as in][]{Labbe:2023}, our spectra imply that at least one-third of the selected objects will be AGN. This number rises to nearly 100\% for the reddest tail of galaxies (F277$-$F444W$>1.6$), and is also $>80\%$ when we apply a compactness criterion and enforce a red power-law continuum. 

An obvious additional question is the possible role of these sources in reionization. Depending on the (poorly understood) number densities and accretion rates of black holes compared to star forming galaxies at the faint-end of the UV LF, they could either have made a minimal contribution \citep[few tens of percent,][]{hassan2018, dayal2020, trebitsch2021,Finkelstein:2022} or dominate the photon budget for reionization \citep[e.g.,][]{Madau:2015, grazian2018}. Furthermore, current models of black hole and galaxy evolution underpredict the number densities of vigorously accreting black holes \citep[e.g.,][]{Furtak:2023nature}. We see this directly in our bolometric luminosity functions, where at the highest luminosities we can only match the number densities if we assume that {\it every} black hole is accreting at its Eddington luminosity. 

It is in principle possible that the occupation fraction approaches unity at high mass. Since we must posit continuous black hole growth at early times to explain the AGN that we are finding \citep[e.g.,][]{Bogdan:2023,Kokorev:2023}, perhaps nature has a way to grow black holes at such high rates. Another possibility is that the black holes outgrow their galaxies dramatically at early times, such that the number densities of available halos is much higher than we assume and the occupation fraction is lowered. We also measure extreme black hole to galaxy ratios in many sources \citep[e.g.,][]{Izumi:2019,Furtak:2023nature}. Finally, we could alleviate the high occupation fractions if we allow sources to exceed their Eddington luminosities. Given the unusual properties of the SEDs of the red AGN, perhaps we are detecting the signature of super-Eddington accretion.

Regardless, we cannot get away from the high number density of black holes at $z>5$. There are a few ways to imagine growing enough black hole mass density this early. Black hole seeds could form heavy (\mbh$\sim 10^4$~\msun) as in direct collapse models \citep{brommloeb2003,loebrasio1994,lodatonatarajan2006,begelman2008,visbaletal2014,Habouzit:2016} or in dense star clusters \citep[e.g.,][]{portegieszwartetal2002, mapelli2016,omukai2008,schleicher2022,Natarajan:2021}. With heavy seeds, it is easier to grow the black hole more rapidly than the galaxy, but most models cannot make high number densities of heavy seeds \citep{dayal2019,Inayoshi:2022}. Alternatively, all of the black holes could start as light seeds \citep{fryeretal2001,brommlarson2004,madaurees2001}, but then grow at super-Eddington rates to make the high mass density of black holes \citep[e.g.,][]{alexandernatarajan2014}. Some super-Eddington accretion is favored by the semi-empirical model TRIDENT \citep{Zhang:2023model}, that uses halo trees to jointly model galaxy and black hole growth. At the same time, detailed magneto hydrodynamical simulations do find viable super-Eddington accretion flows \citep[e.g.,][]{Jiang:2019}. Super-eddington accretion may also explain some of the SED properties of the compact red sources, in particular the apparently low X-ray luminosities \citep{Matthee:2023,Furtak:2023nature} and even possibly the red continuum if the inner accretion flow grows optically thick but leaves an outer disk. Finally, super-Eddington accretion would alleviate the tension with our ``maximal'' model, but producing more luminosity for a given black hole.

One other complication may come from mergers. At low redshifts, reddened broad-line AGN seem to preferentially reside in merging hosts \citep[e.g.,][]{Urrutia:2008}. If the red sources presented here were also dominated by merging hosts, then perhaps the merging system may be harder to detect than an undisturbed galaxy, due to variable extinction and perhaps a significant low surface brightness component. There could also be two AGN powering the objects in some cases in principle. Notionally, the number densities of major mergers at $z \sim 6$ could be high enough to match the number density of the compact red sources. Taking an empirically motivated merger timescale of $\sim 0.5$~Gyr and a merger volume density of $\sim 3-5 \times 10^{-5}$~Mpc$^{-3}$~Gyr$^{-1}$ for $M_* \approx 10^9-10^{10}$~\msun\ galaxies from empirical halo modeling \citep{Oleary:2021}, we estimate $\sim 2 \times 10^{-5}$~Mpc$^{-3}$ mergers at $5<z<6$. Plausibly then there could be a relationship between the AGN triggering, merging, and the observed reddening. As was seen in \citet{Matthee:2023}, the compact red sources also appear to be clustered with each other. \citet{Fujimoto:2023overdense} highlights a potential overdensity hinted from a compact red AGN and a UV-bright object found together in the same giant ionized bubble with a radius of $7.69\pm0.18$ proper Mpc at $z=8.5$. Perhaps this excess clustering could be related to a merger origin for these sources.

There are many additional puzzles raised by the 'little red dots', including their apparent clustering, their unique SEDs (characteristic red optical continuum, an additional UV component, and lack of X-ray emission), and their apparent lack of a significant host galaxy component. These red broad-line AGN apparently constitute a sizable $10-20\%$ fraction of broad-line AGN at $z>5$, as well as a sizable fraction of red galaxies at the same epoch. They are an important part of the story of black hole growth at early times.

\section*{Acknowledgments}

J.E.G. and A.D.G acknowledge support from NSF/AAG grant\# 1007094, and J.E.G. also acknowledges support from  NSF/AAG grant \# 1007052. A.Z. acknowledges support by Grant No.~2020750 from the United States-Israel Binational Science Foundation (BSF) and Grant No.~2109066 from the United States National Science Foundation (NSF), and by the Ministry of Science \& Technology of Israel. The Cosmic Dawn Center is funded by the Danish National Research Foundation (DNRF) under grant \#140. This work has received funding from the Swiss State Secretariat for Education, Research and Innovation (SERI) under contract number MB22.00072, as well as from the Swiss National Science Foundation (SNSF) through project grant 200020\_207349. PD acknowledges support from the NWO grant 016.VIDI.189.162 (``ODIN") and from the European Commission's and University of Groningen's CO-FUND Rosalind Franklin program. K.G. and T.N. acknowledge support from Australian Research Council Laureate Fellowship FL180100060. H.A. and IC acknowledge support from CNES, focused on the \jwst\ mission, and the Programme National Cosmology and Galaxies (PNCG) of CNRS/INSU with INP and IN2P3, co-funded by CEA and CNES. RPN acknowledges funding from \jwst\ programs GO-1933 and GO-2279. Support for this work was provided by NASA through the NASA Hubble Fellowship grant HST-HF2-51515.001-A awarded by the Space Telescope Science Institute, which is operated by the Association of Universities for Research in Astronomy, Incorporated, under NASA contract NAS5-26555. The research of CCW is supported by NOIRLab, which is managed by the Association of Universities for Research in Astronomy (AURA) under a cooperative agreement with the National Science Foundation. 
A.J.B.\ acknowledges funding support from NASA/ADAP grant 21-ADAP21-0187.
Support for this work was provided by The Brinson Foundation through a Brinson Prize Fellowship grant.
RPN acknowledges support for this work provided by NASA through the NASA Hubble Fellowship grant HST-HF2-51515.001-A awarded by the Space Telescope Science Institute, which is operated by the Association of Universities for Research in Astronomy, Incorporated, under NASA contract NAS5-26555.  CP thanks Marsha and Ralph Schilling for generous support of this research.


\end{document}